\documentclass[preprint]{aastex}

\slugcomment{The Astronomical Journal, in press}

\begin{document}
\title{The Environments of a Complete, Moderate-Redshift Sample of {\it FIRST} Bent-Double Radio Sources}
\author{E. L. Blanton\altaffilmark{1,4}, 
M. D. Gregg\altaffilmark{2,3}, 
D. J. Helfand\altaffilmark{4}, 
R. H. Becker\altaffilmark{2,3}, and 
K. M. Leighly\altaffilmark{4,5}}

\altaffiltext{1}{Department of Astronomy, University of Virginia,
P. O. Box 3818, Charlottesville, VA 22903-0818;
eblanton@virginia.edu}

\altaffiltext{2}{Physics Department, University of California, Davis, CA 95616}

\altaffiltext{3}{Institute of Geophysics \& Planetary Physics, Lawrence 
Livermore National Laboratory, Livermore, CA 94550; gregg@igpp.ucllnl.org,
bob@igpp.ucllnl.org}

\altaffiltext{4}{Columbia Astrophysics Laboratory, 550 West 120th St., 
New York, NY  10027; djh@astro.columbia.edu}

\altaffiltext{5}{current address:  University of Oklahoma, Department of Physics and Astronomy,
440 W. Brooks St., Norman, OK  73019; leighly@mail.nhn.ou.edu}

\begin{abstract}
We present an optical spectroscopic and imaging study of the environments of a complete sample of
moderate-redshift, bent-double radio sources.  More than half of the forty radio galaxies in the sample are
associated with Abell richness class 0 or greater clusters at $z<0.4$.  Most of the
remaining objects are associated with groups, although a few appear to be hosted by
nearly isolated elliptical galaxies.  For the
bent doubles appearing in
poor environments, either dense gas must be associated with the systems 
to provide the ram pressure to bend the lobes, or alternative bending 
mechanisms will have to be invoked to explain the radio morphologies.
Correlation with the $ROSAT$ All Sky Survey Bright and Faint Source Catalogs
reveals the majority of the $z<0.2$ objects in our sample that we 
classify optically as clusters are also X-ray sources.

\end{abstract}

\keywords{
galaxies:clusters:general ---
galaxies: elliptical and lenticular, cD ---
galaxies: intergalactic medium ---
radio continuum: galaxies ---
X-rays: galaxies}

\section{ Introduction}

With increasing redshift, optical and X-ray survey techniques for finding
galaxy clusters lose their effectiveness and a targeted approach is warranted.
In particular, radio galaxies can be used as cluster signposts. Dickinson (1996) has identified clusters around powerful 3C sources
at redshifts greater than 1.2, and Deltorn et al. (1997) discovered a cluster of
galaxies around 3CR 184 at $z = 0.996$.  Clusters of varying richness have
also been seen around the radio galaxies studied by Hill \& Lilly (1991 -- 
hereafter H\&L91) and by
Zirbel (1997 -- hereafter Z97), among others.  However, some kinds of radio 
galaxies are more often associated with clusters than others, and are 
therefore more efficient tracers of these high-density environments.

Over two decades ago, Fanaroff \& Riley (1974) discovered a correlation
between the power of an extended radio source and its
morphology, and divided all such sources into two classes.  
FR I radio galaxies
typically have powers less than $5\times10^{25}~\rm{W~Hz^{-1}}$ at 1440 MHz
(assuming a power law spectrum with $\alpha = 0.8$), bright radio cores, and
lobes fading toward the edges.  
FR II's have $\rm{P_{1440}>5\times10^{25}~W~Hz^{-1}}$ with dim or absent 
cores and edge-brightened lobes.  More recent work has shown that the dividing
line in power between FR I's and II's is a function of optical luminosity
(Ledlow \& Owen 1996; Owen \& White 1991).
 
FR I and II radio sources are associated with host galaxies located in different
environments on the megaparsec scale (H\&L91; Z97).  Elliptical galaxies with a 
wide range of optical magnitudes can play host to both FR I and II sources (Ledlow 
\& Owen 1995a; Ledlow, Owen, \& Eilek 2000).  The more powerful FR I galaxies (those
near the FR I/II break such as the wide-angle and narrow-angle tail sources 
discussed below) are often associated with cD or D galaxies found in rich groups or 
clusters at all redshifts up to z $\approx$ 0.5, the limit of current studies.  FR 
II sources are usually found in poor groups at low-z, although at higher
redshifts ($z \approx 0.5$), groups surrounding FR II's appear to
be richer by a factor of $\sim 2$.  The hosts of some FR II's are quasars.
 
A source's radio morphology may become distorted as a consequence of 
the relative motion between the host
galaxy and the surrounding intracluster medium (ICM) or via interaction with a neighboring galaxy.  Common
distortions in FR I sources include wide-angle tail (WAT) and narrow-angle tail
(NAT) morphologies which have large and small opening angles, respectively.
The classical explanation for
the bending of the lobes is the ram pressure exerted by the ICM as a galaxy
with a significant peculiar velocity moves through a cluster (Owen \& Rudnick 1976).
  An alternate
explanation (Burns et al. 1996; Roettiger, Burns, \& Loken 1996; Burns et al. 1993) 
has the ICM in motion rather than the galaxy, since WATs are often associated with cD
galaxies which have small peculiar velocities relative to the immediately
surrounding cluster.  In this scenario, the
ICM is set in motion by the merging of clusters or smaller sub-clumps; evidence
for this view is found in the alignment of X-ray emission contours with the bending of the lobes
(G\'omez et al. 1997).  FR II sources may also become 
distorted; sometimes the distortions appear symmetric and may result from ICM
interaction, whereas at other times the sources have an asymmetric appearance
most likely caused by interaction with a neighboring galaxy (Rector, Stocke,
\& Ellingson 1995).
 
A VLA survey of nearby Abell clusters (Zhao et al. 1989; Ledlow \& Owen
1995a, 1995b, 1996) found that 95\% of the radio sources in the clusters were
type FR I.
In a radio continuum survey of distant ($z = 0.3 - 0.8$) X-ray-selected clusters
from the $Einstein$ Medium Sensitivity Survey, 
Stocke et al. (1999) found that $all$ cluster radio
sources were of type FR I, and most exhibited WAT
or NAT morphologies.  They found no difference in the radio luminosity 
function at these redshifts compared to that seen nearby, and concluded that
there is no evidence for evolution in the population of rich cluster
radio galaxies between $z \sim 0.5$ and the present epoch.
To explain the absence of FR II sources in these clusters, Stocke et al.
hypothesize that when FR II sources seem to appear in cluster environments
(as in H\&L91 and Z97), the clusters they appear
in may contain a lower density intracluster medium.  By definition, all of the
clusters in  the Stocke et al. sample have dense ICM because they were selected
as bright X-ray sources.

FR I bent-double radio sources (WATs and NATs) then, appear to 
trace out high-density regions in the Universe.  These may appear as
clusters, groups, or even 'fossil-groups' -- large elliptical galaxies that have cannibalized their neighbors leaving behind the dense gas once
associated with the doomed cluster or group (Ponman et al. 1994; Vikhlinin
et al. 1999; Romer et al. 2000).

Previous authors have surveyed the radio properties of clusters selected at
optical or X-ray wavelengths (e.g. Ledlow \&
Owen 1995a, 1995b, 1996; Stocke et al. 1999) or have studied the environments of
mostly high-power FR I and II radio sources (e.g. Hill \& Lilly 1991; Zirbel
1997).  Here we examine the environments of a morphologically selected sample
of radio sources.
We derived our sample of bent-double radio galaxies from the VLA {\it FIRST} 
survey (Faint Images of the Radio Sky at Twenty-cm; Becker, White, \& Helfand
1995)
with the initial goal of using them as tracers of distant clusters of galaxies
(Blanton et al. 2000).  We found that the bent doubles, combined with optical
selection criteria, had an 80\% success rate at finding distant clusters
(up to $z=0.84$).  Here, we present a complete, low- to moderate-redshift 
sub-sample
of the bent-double sources in order to study their environments, as well as to
quantify their success rate at
finding clusters.  In addition, we search for correlations between the radio
and optical properties of the sources and their surroundings to determine 
whether any
radio properties point toward cluster association.  Finally, we correlate
our sample with the ROSAT All Sky Survey.

We use $H_{\circ}=50~\rm{km~s^{-1}~Mpc^{-1}}$ and, except as otherwise noted,
$q_{\circ}=0.5$ throughout this paper.

\section{The Sample}
Our sample of radio galaxies was selected from the {\it FIRST} survey 
currently
being undertaken at the VLA using the B configuration (Becker, White, \&
Helfand 1995).  The survey
is projected to cover the 10,000 deg$^2$ of the North Galactic Cap plus a
supplementary region between $-10^{\circ}$ and $+2^{\circ}$ Dec in the range
 21.5$<$R.A.$<$3.5 h.  The peak
flux density threshold is 1.0 mJy at 20 cm, the angular resolution is 
5$^{\prime\prime}$, and the astrometric accuracy for all detected sources is
better than $1^{\prime\prime}$.  In
the $\sim$ 3000 deg$^2$ region surveyed as of the April 1997 catalog release, 
{\it FIRST} detected approximately
270,000 radio sources of which $\sim$ 32,000 are double or multiple 
sources with separations less than $60^{\prime\prime}$  
(as of this writing, {\it FIRST} has covered 8000 deg$^2$ and detected 722,000 
sources).  
From the April 1997 catalog, we selected
384 sources which exhibit the bent-double morphology.  The peak
and integrated 20 cm flux densities of these sources are typically 
a few mJy and
10 - 100 mJy, respectively.  

The sample was constructed by visually inspecting grayscale plots of all
{\it FIRST}
sources having more than one component within a circle of radius 60$^{\prime\prime}$.
All images were examined independently by three of us (ELB, RHB and DJH);
unambiguous distortion of the radio lobes was required. 
If at least two out of three examiners selected a 
source,
it was included in the sample.  Objects chosen by only one of us
were re-examined, and a few of these were added to the sample.
The sample contains a large fraction of FR I radio 
galaxies, as these are more frequently found to be distorted than are FR II's 
which most often 
appear as collinear classical doubles.

Our sample of 384 objects consists of 50\% FR I sources, 25\% FR II sources, and 25
\% FR I/II (sources with intermediate 
morphologies).  
Thirty-three (8.5\%) of these sources are within 5 arcmin of the center of
an Abell 
cluster.  We selected and classified these before knowing of their cluster 
associations; of these 33, we classified 80$\%$ as FR I, 10$\%$ as FR II, 
and 10$\%$ as FR 
I/II.  The bent-double sources are easily recognizable by {\it FIRST} to $z > 1$,
providing us with a good opportunity to find new moderate- to
high-redshift clusters.
Some of the sources classified as FR I/II and II may be FR I's that have 
had extended emission resolved out by {\it FIRST} and/or have had their 
apparent morphologies altered by
1/$(1+z)^4$ surface brightness dimming.  However, the high surface 
brightnesses 
and small
angular extents of the lobe hot spots which characterize FR II's make
it unlikely that an FR II would be misclassified as an FR I.
Thus, even after allowing for some 
classification errors, the majority of our sources are of type FR I.
The most common morphology in our sample is the wide-angle-tail (WAT) radio 
source.  

\subsection{Defining a Complete Sample}
The complete sample studied here was chosen to be area- and magnitude-limited.
It contains all bent-double sources from our sample of 384 within the range $\rm{8h < RA < 14h}$ and
$\rm{+22^{\circ} < Dec < +31^{\circ}}$, and
with R-magnitudes brighter than m$_{R} = 19$ for the optical
counterparts. There are 99 bent doubles in
the RA and Dec ranges listed above, of which 45 have host magnitudes above our chosen limit.  
Two objects were eliminated because their spectra indicated
that they were stars (at least one of these was an obvious misidentification);
in addition,
one object was eliminated because we did not obtain a redshift for it although
it fell just inside the magnitude limit, one object had a spectrum 
but no image taken, while one was not observed at all due to observing-time 
constraints.  
Our final sample includes 40 objects.
A log of the optical and spectroscopic observations is given
in Table 1.
Postage-stamp grayscale images of the radio sources are
displayed in Figure 1, and the properties of the bent doubles are presented 
in Table 2.
Column (1) lists the identification number, and columns (2) and
(3) give the source position, which is either the position of the radio
core if there is one, or the position of the optical identification. Column (4) lists the  redshift, column
(5) contains the total R-magnitude, 
column (6) lists the integrated flux density taken from the NRAO/VLA Sky Survey
({\it NVSS}, Condon et al. 1996), since
{\it FIRST} may resolve out some of the extended emission. 
Column (7) lists the ratio of {\it NVSS} to {\it FIRST} flux,
column (8) lists the power
at 1440 MHz (assuming a power law spectrum with $\alpha=0.8$), column (9)
gives the opening angle of the bent-double source, column (10) lists the
total unbent angular size of the source -- from the edge of one lobe to
the core, and out to the edge of the other lobe. Column (11) lists the 
corresponding linear size, while column (12) gives our estimate of the Fanaroff
\& Riley (1974) classification (where a class of I/II means it was difficult
to determine),
and column (13) gives comments including
known cluster associations.  Unless otherwise stated in column (13), the optical
identifications for all of these appear to be elliptical galaxies.

\section{ Optical Observations and Initial Analysis}
\subsection{ Images}
All of the objects were observed
at the MDM Observatory with the 2.4m telescope and the Echelle CCD
with the exception of three sources (\#082, 236, and 237) for which imaging 
information was taken from the literature. 
The observations were accomplished during three February observing runs, in
1998, 1999, and 2000.  The observing date and
exposure time for each object are listed in Table 1.  All of the bent-double
fields were observed through an R-band filter.  The 1998
run used a 2-inch R$_{KPNO}$ filter and the
 1999 and 2000 runs employed the 4-inch R$_{MDM}$ filter.  The Echelle
CCD has an area of 2048$\times$2048 pixels and an image scale of $0.^{\prime\prime}$275 arcsec/pixel, yielding a total field of view of 9.4 arcmin$^2$.
This f.o.v. was reduced to 8.0 arcmin$^2$ for the 1998 run (from 
which 
we include observations for only three of the bent doubles)
because the filter was not large enough to cover the entire CCD.

For the 1998 and 1999 runs, three exposures were taken for each field, dithering in-between so that
chip defects would average out. For the 2000 run, one long exposure was taken
for each bent double.  Standard stars were
observed at least three times during the night.
The images were corrected for bias, overscan, and flatfield effects in
a standard manner using IRAF\footnote{IRAF (Image Reduction and Analysis Facility) is distributed by the National
Optical Astronomy Observatories, which are operated by the Association of
Universities for Research in Astronomy, Inc., under cooperative agreement
with the National Science Foundation}.
The separate exposures taken for each bent-double field were aligned and
average-combined to remove cosmic rays and increase the 
signal-to-noise ratio.  Those fields with only one exposure were cleaned for
cosmic rays (defined as pixels above a certain flux ratio as compared with neighboring
pixels) using the IRAF task 'cosmicrays.'

Observations were taken under photometric conditions.  Standard stars from
the Landolt (1992) catalog were used for calibrations.
Color, extinction, and zero-point terms were derived each night.
The average (mean) color term
derived for the 1999 observations was $T_{r} = -0.07\pm0.02$ and the
average extinction term was $k_{r} = -0.15\pm0.07$.  For the 2000 
observations, we found $T_{r}=-0.07$ and $k_{r}=-0.10$.  For the one-night
 1998 run, the derived color term was negligible ($T_{r}=0.01$) and,
because of standard star limitations, we were unable to derive a reliable 
extinction term; standard stars observed at the same airmasses
as the objects were used to derive the photometry.  The photometry for 
objects observed during both the 1998 and 2000 runs is consistent: the 1998
magnitudes are higher by $0.08\pm0.08$ mag.  

A catalog of objects was created for each bent-double field containing 
all sources detected at least 3$\sigma$ above the sky level. 
Total magnitudes were measured using FOCAS (Faint Object Classification 
and Analysis System -- Valdes 1983) and are listed in Table 2. The area used for determining
the ``total'' magnitude was found by increasing each object's isophotal
area by a factor of two. FOCAS missed some of the flux for very bright,
extended objects and the magnitudes listed for these have possible errors
as high as $\pm0.2$ mag. As we obtained only R-band data, an estimate of the (V-R) color of an elliptical
galaxy at the redshift of the radio host was taken from the no-evolution
models of Coleman, Wu, and Weedman (1980).

\subsection{ Spectroscopy}
A log of the spectroscopic observations is included in Table 1.  Most of
the observations were performed at the MDM 2.4m telescope with the MkIII
spectrograph and the Echelle CCD.  Redshifts for three objects (\#82, 236, and
237) were taken from the literature, while spectra for two objects are from
Lick Observatory, one is from the KPNO 4m telescope, and one was observed
with the Keck II telescope (Blanton et al. 2000).  Spectra for an 
additional four objects were obtained at MDM in December 1998 and were 
kindly provided by J. Halpern.
The majority of the  spectroscopic
observations at MDM were completed during two runs -- February 1998 and
1999.
They were done in longslit mode,
with slit widths of 1.$^{\prime\prime}2$ (February 1998) and 1$^{\prime\prime}.68$ (February 1999).  Both runs used
the 600 lines/mm grism with a central wavelength of 5800 $\rm{\AA}$.  This setup
gave a dispersion of 2.27 $\rm{\AA}$/pixel. High signal-to-noise ratio data were 
obtained in the wavelength range 4500 -- 8000 $\rm{\AA}$.  The spectral resolution is
approximately 10 $\rm{\AA}$.

The spectroscopic frames were reduced following standard IRAF procedures.  They were
corrected for bias, overscan, and flatfield effects.    
When only one exposure of a field was available 'cosmicrays'
was run to clean the field.
If two exposures were taken, they were averaged with cosmic rays rejected.
A spectrum of the
bent-double radio source was extracted in each case, and was wavelength- and
flux-calibrated.  The wavelength calibrations were derived using comparison
lamp observations taken at the beginning and/or end of each night. The
data were flux-calibrated using observations of KPNO spectrophotometric
standard stars.  

All of the spectra, excepting those of the quasars (\#115, 189, and 249), 
had absorption lines consistent with old stellar populations that would
be expected in elliptical galaxies (consistent with their morphologies). 
Four (11\%) of the non-quasar spectra also had emission lines, typically of
[OII] $\lambda3727$, H$\beta$ $\lambda4861$, and [OIII] $\lambda4959$ and
$\lambda5007$.

Redshifts were obtained by
performing a Fourier cross-correlation between each galaxy's spectrum and
a template (M32 shifted to the rest-frame), using the task FXCOR in IRAF.
Errors were computed in FXCOR and represent the fitted peak height and the
asymmetric noise, or ``r-value'' (Tonry \& Davis 1979), and are typically
100 -- 200 km s$^{-1}$.  Each of
the three quasars had only one broad, bright, emission line, which we
assume is Mg II, since other prominent lines would not be expected to appear alone over
the wavelength coverage of our spectra.  Redshifts for all objects in the
sample are listed in Table 2.

\section{Radio Properties}
The radio properties of the sources are also listed in Table 2.
Most of the sources have powers typical of WAT radio sources, within a factor
of three of 5 $\times 10^{25}$ W Hz$^{-1}$ at 1440 MHz (O'Donoghue et al. 1993).
Of the 40 sources, we classify 27 (68\%) as FR I, 8 (20\%) as FR II, and
5 (12\%) as FR I/II.  This is similar to the distribution of types for
the whole 384-object bent-double sample.  The higher fraction
of FR I's in our low-to-moderate redshift complete sample is at least
partly due to surface brightness 
dimming of high-$z$ sources; the extended lobe 
emission may not be detected in FR I's at high-$z$, leading to erroneous 
classification as FR II's, and producing a higher fraction of FR II's in the parent sample.  

The angular resolution of the {\it FIRST} survey is necessary to recognize the bent
double morphology and to determine unambiguously the optical counterpart.
However, the $5^{\prime\prime}$ angular resolution of the {\it FIRST} survey resolves out some of the extended emission. Thus, we use the {\it NVSS} survey (angular resolution $\approx 45^{\prime\prime}$) in reporting the source 
integrated flux densities in Table 2.  The mean
ratio of {\it NVSS} to {\it FIRST} flux density for the individual sources in the complete sample
is 1.3, with a range in
values from 1.0 to 3.1.

\section{ Richness of the Bent-Double Fields}
A catalog of objects detected at least 3$\sigma$ above the sky was created
in FOCAS for each bent-double field.  
Total magnitudes were 
determined by measuring the flux within twice each object's isophotal
area.  Sky values were determined locally.  To improve sky subtraction and avoid
subtracting off flux from an object's extended envelope, we
increased the distance of the sky buffer (the distance from the edge of an object's
isophote to the edge of the annulus where sky is measured) from the FOCAS default value
to be large enough that magnitudes no longer changed as the buffer distance was 
increased.

Richness values for the bent-double fields were estimated following 
the method used by Allington-Smith
et al. (1993) and Z97.
The richness statistic $N_{0.5}^{-19}$ measures
galaxies within a 0.5 Mpc radius of the radio galaxy, and with
absolute magnitudes brighter than $M_{V}=-19$ (which is $M_{V}^{*}+2.9$).
This
is similar to an Abell-type richness measurement, although Abell (1958, 1989)
used the
magnitude range $m_{3}$ to $m_{3} + 2$, where $m_{3}$ is the third brightest
cluster member.  In all but very nearby clusters, and without spectroscopy
for at least several members, the third brightest member is difficult to
identify because of the superposition of foreground and background galaxies on
the plane of the cluster.  In addition, the $N_{0.5}^{-19}$ statistic has the
virtue of
measuring the same absolute magnitude range in each potential cluster. 

To transform $R$-magnitudes to absolute $V$-magnitudes, an elliptical galaxy
spectrum was assumed with a rest frame color $(V-R)=0.9$, and K-corrections
were taken from Coleman, Wu, \& Weedman (1980).  No evolution correction was
applied.  Galactic extinction values were taken from Schlegel, Finkbeiner,
\& Davis (1998).

To determine the density of galaxies surrounding each radio source, our raw
richness counts needed to be corrected for unrelated field galaxy
counts; we call these ``background counts.''  Ideally, this would be achieved
by counting background galaxies directly on the frames.  However, particularly
for the lowest-$z$ fields, our $9.4 \times 9.4$ arcmin images will contain
associated cluster or group galaxies even at the edges of the frames and the
background would therefore be overestimated.  The other possibility is to
use published background counts of galaxies from deep surveys.  We have used
a combination of the two methods.  

A plot of the number of galaxies and stars (per 0.1 mag per deg$^2$) catalogued on all 33 frames for which we determined a richness value
is shown in Figure 2.  From this plot, it is evident that on average, our
frames are complete to approximately $m_{R}=24$ where the star and
galaxy counts both suddenly plummet.  In addition the star/galaxy 
classification in FOCAS breaks down at $m_{R} \approx 21.5 - 22$ for our
fields.   Results from the deep survey of Metcalfe et al. (1991) indicate 
that star counts make up 
approximately 10\% of the total counts at $m_{R} = 21$ and less than 2\% at
$m_{R} = 23$.  As seen in Figure 2, the fraction of stars at these
magnitudes determined by FOCAS for our frames is much higher; it is obvious
that, at faint magnitudes ($m_{r} > 21.5$ or 22), many galaxies are being 
erroneously classified as stars.  In order to account for this, we include
objects classified as stars, and with magnitudes fainter than $m_{R} = 21.5$
and down to the $m_{R}$ corresponding to $M_{V}=-19$, in our richness 
measurements.

Only the higher redshift
bent doubles will be affected by this misclassification, because it is only at relatively
high $z$ that $M_{V}=-19$ corresponds to a sufficiently faint apparent 
magnitude that misclassified objects affect the richness statistic.
The lowest $z$ object affected by this is \#177 at
$z = 0.256$.  For the bent doubles at higher $z$, we use background counts 
determined directly
from the image frames and also include objects classified as stars, with
magnitudes in the range $m_{R} > 21.5$ to the $m_{R}$ corresponding to 
$M_{V}=-19$, in the background counts.  For the lower redshift objects, where
we expect the most contamination from group or cluster galaxies to the frame
background counts, and where
the apparent magnitude corresponding to $M_{V}=-19$ remains relatively bright,
we use background counts from Metcalfe et al. (1991).  Metcalfe et al. found
log(N) = $0.37m_{R} - 4.51$, where N is the number of galaxies per deg$^2$ per
0.5 mag.  These counts agree well with several other surveys including those of
Steidel \& Hamilton (1993) and Lilly et al. (1991).  

As a check, we determined
the background counts for all fields using both methods: the Metcalfe et al.
relation, and the counts in an annulus surrounding each radio source.  
For the lower-$z$ bent doubles ($z < 0.256$),
the frame counts are higher, by a mean of $4.0\pm0.8$ (error in the mean).
This is expected because of the contribution from cluster or group galaxies 
to the frame background counts.  For the higher-$z$ ($z > 0.256$) fields, the 
two methods agree well.  The mean difference
between the frame and Metcalfe et al. background counts is $-2.3\pm2.0$; 
rejecting one extreme outlier gives a mean of $-0.6\pm1.4$.

Table 3 lists the results of the richness measurements.  Column (1) 
lists the bent-double's id number,
column (2) gives the rank of the radio galaxy in its potential cluster
(it is the first ranked, or BCG, in all but 8 of the 40 cases),
column (3) gives the absolute V-magnitude of the radio source host galaxy, and
column (4) lists $m_{R,-19}$, defined as the apparent R-magnitude corresponding
to $M_{V} = -19$ at the redshift of the bent double.
Column (5) lists the raw $N_{0.5}^{-19}$ counts, column (6) lists the field
galaxy counts calculated from Metcalfe et al. (1991) if $m_{R,-19} < 21.5$, or
determined directly on the frame if $m_{R,-19} > 21.5$, column (7) lists the 
corrected
richness values (raw - background counts) along with Poisson errors, column (8) gives our estimate
of the Abell richness class of the cluster based on Z97's 
transformation
between $N_{0.5}$ and Abell class, and column (9) tells of known cluster
or group associations.  To estimate the Abell richness class (column 8), we used the transformation $N_{Abell}=
2.7(N_{0.5}^{-19})^{0.9}$, from Z97 with $q_{\circ}=0$, and Abell's (1958) 
conversion between $N_{Abell}$ and richness class.  Using this relation,
inclusion in Abell's catalog at the lowest richness level (richness class
0) requires $N_{Abell} = 30$ or $N_{0.5}^{-19}=14.5$ ($q_{\circ}=0$).
Since we have
assumed $q_{\circ}=0.5$ in this paper, we transform our richness statistic to
the value it would have with $q_{\circ}=0$ before estimating the Abell class.
For $q_{\circ}=0.5$ the sampling radius in which galaxies are counted 
increases relative to that calculated for $q_{\circ}=0$.
This combined with a decline in the number of galaxies from the center of a
cluster that goes as $N_{gal}\propto r^{0.8}$ (Allington-Smith et al. 1993),
means that $N_{0.5}^{-19}$ with $q_{\circ}=0.5$ is higher than
$N_{0.5}^{-19}$ with $q_{\circ}=0$ by 8\%, 6\%, 4\%, and 2\% for 
$z = 0.4, 0.3, 0.2,$ and 0.1, respectively.

We did not make a richness measurement for object
\#082 because the image was not large enough to include a 0.5
Mpc radius around the radio source (this source is at very low-$z$).
In any event, this source has been studied previously, and is known
to be associated with a poor group (Stocke \& Burns 1987).  In addition,
we did not make our own richness measurements for objects \#236 and 237, since
they reside in the Coma cluster and sufficient data exist for them.
We did not calculate richness values in the fields surrounding the three
quasars as our images were not deep enough to measure galaxies at the
redshifts of the quasars.

Four of our fields for which we measured the richness class contain the
clusters Abell 824, 876, 1258, and 1587.
A comparison of the richness class given by Abell (1958, 1989)
shows that our quantitative estimates agree with Abell's to within $\pm1$ Abell class:
for two of the clusters, our classification agrees, for another, our
estimate is one class higher than Abell's, and the remaining one falls just
short of our cluster cutoff -- we would classify it as a rich group.

Converting our richness measurements, using $q_{\circ}=0$ for a direct 
comparison to Z97's results, we found a mean $N_{0.5}^{-19}=16.7\pm{2.2}$ 
($17.4\pm{2.3}$ with $q_{\circ}=0.5$) for all of the sources in our complete sample.
Evaluating FR I and II objects separately (excluding those classified as
FR I/II, and approximating $N_{0.5}^{-19}=8$ for object \#082 and
$N_{0.5}^{-19}=55$ for object \#236), we find a mean 
$N_{0.5}^{-19}=16.2\pm{2.4}$ ($16.8\pm{2.5}$ with $q_{\circ}=0.5$) 
for the 26 FR I sources and $N_{0.5}^{-19}=12.8\pm{5.8}$ ($13.8\pm{6.1}$ 
with $q_{\circ}=0.5$) for the six non-quasar FR II sources.  Z97 found a mean value 
of $N_{0.5}^{-19}=8.6\pm{0.9}$ for all sources in her sample, with 
$N_{0.5}^{-19}=14.3\pm{2.0}$ for FR I's and $N_{0.5}^{-19}=8.7\pm{1.2}$ for 
FR II's.  

We find no significant difference in the richness values for the environments 
surrounding the FR I and FR II sources in our sample.  However, the number of 
FR II sources in our sample is very small (six).
The mean richness value for our sample of mostly FR I sources is higher than
the mean richness value of the Z97 sample, which contains mostly FR II sources,
by a factor of approximately two.
When we compare the mean richnesses of the environments surrounding the FR I's
in our sample and Z97's sample, however, we find no significant difference.
Specific morphologies (i.e., bent or collinear) were not noted in Z97.  However,
since the FR I sources in the Z97 sample have powers near or above the FR I/II
break, it is likely that many of them are WATs.  In fact, the mean power for
the FR I's in the Z97 sample, converting to 1440 MHz assuming
$P\propto\nu^{-\alpha}$ with $\alpha=0.8$,
is an order of magnitude higher than the mean power of the FR I's in our 
sample:
mean $P_{1440}=1.5\times10^{26}$ W Hz$^{-1}$ with a range of 
$P_{1440}=3.6\times10^{25}$ W Hz$^{-1}$ to $3.6\times10^{27}$ W Hz$^{-1}$ 
(Z97) vs.
mean $P_{1440}=1.4\times10^{25}$ W Hz$^{-1}$ with a range of 
$P_{1440}=6.0\times10^{23}$ W Hz$^{-1}$ to $5.9\times10^{25}$ W Hz$^{-1}$ 
(our sample).
It is unclear if there is any difference in the gas content of the
source environments in the two samples -- in particular, whether our sample, 
which was specifically chosen by radio morphology, contains sources that reside in 
environments that harbor denser gas more often than powerful FR I 
environments in general.  X-ray observations are necessary to decide this.

There is an approximately fifty-fifty chance of finding a cluster when optically following
up on a bent-double radio source.  From our richness values, we find that 54\% of
the bent doubles in the complete sample are found in clusters of Abell
richness class 0 or greater.  Figure 3a shows an example of
a bent double in a rich environment.  
In addition, 46\% of the objects are found in groups or the field, suggesting
either that dense gas can be found in groups,
some of them rather poor, or that there are alternative bending mechanisms
distorting the
radio lobes.  An example of a source found in a poor environment is displayed in Figure 3b.

Venkatesan et al. (1994) claim that narrow-angle-tail (NAT) sources may be found in poor environments because
it is the {\it local} galaxy environment, rather than the global cluster environment, that produces the dominant effect on the radio morphology.  In their model,
a poor cluster ICM combined with high-velocity, infalling galaxies can bend
radio lobes.  Such high velocities would be possible in poor clusters in the 
early stages of collapse. Burns et al. (1994) found clumped X-ray emission around radio sources in
Abell clusters.  While some of the X-rays may come from the AGN powering
the extended radio source, the emission is often extended.
Their study concludes that local clumped gas may bend
radio lobes, and that this gas is a signature of a recent cluster-cluster
(or group) merger.  Wide-angle-tail (WAT) radio sources, then, may be 
pointers to clusters in formation, or those that have recently undergone
mergers.  This is consistent with the observation that WAT clusters generally
do not have cooling flows, which are found in relaxed clusters.

Another theory, put forth by Stocke \& Burns (1987) is that the low dynamical
pressure resulting from gas associated with a group of galaxies allows
radio jets to expand more quickly.  This, in turn, lowers the density and
pressure of the jets, allowing them to be bent more easily.

In X-ray surveys, a new class of objects called ``fossil groups'' has recently
been discovered (Ponman et al.\ 1994; Vikhlinin et al.\ 1999; 
Romer et al.\ 2000).  These objects display extended
emission in the X-ray, and their optical counterparts are single, large, bright
elliptical galaxies that appear to reside in poor environments.  They may
represent groups of galaxies that have been ``cannibalized'' by the bright
elliptical, leaving behind only the extended X-ray gas.  The dense gas associated
with these systems may represent a significant mass component to the Universe
that would be missed by optical surveys.

\section{ The Distributions of Source Properties and their Environments}
We discuss here the statistical properties of the complete sample; in all
that follows, we exclude the three quasars (all high-power FR II objects)
from further consideration.

All but eight of the 37 host galaxies are
brightest cluster/group galaxies.  Furthermore, the scatter in 
host absolute magnitude is small, with all but eight of the objects lying
within $\pm 0.5$ mag of the median value of M$_R = -22.8$ (see Figure 4).
Indeed,
the apparent magnitude distribution (Figure 5) is very similar to a plot of R-mag vs. $z$
for the no-evolution model of an elliptical galaxy (Coleman, Wu, \& Weedman
1980).  This is consistent with the results of Ledlow \& Owen (1995b) who found that the host elliptical galaxies of
FR I radio sources are indistinguishable from elliptical galaxies with
no radio emission.  They conclude that, given the lifetimes of radio sources
($\le 1.4 \times 10^9$ yrs), all elliptical galaxies may at some point contain powerful radio sources.

The distribution of radio power, plotted as a function of redshift, is 
displayed in Figure 6. Most of the sources lie below the canonical FRI/FRII 
dividing line and three-quarters of the objects lie at least a factor of two 
below this boundary. Apart from the expected lack of rare, high-power
sources in the small local volume and the lower envelope produced by the 
survey's radio flux density threshold, no trends with $z$ are apparent. There
is no correlation between radio power and host galaxy magnitude, with radio
power ranging over two orders of magnitude for a given optical luminosity.
This is also consistent with the findings of Ledlow \& Owen (1995a; 1996).

There is also no correlation seen between the redshift of the radio source and the richness of
its associated cluster (Figure 7).  Out to $z\sim0.5$, the cluster
environments are just as rich as they are at present day.  This is consistent
with the results of H\&L91 for their subsample of FR I objects.
Cluster richness is also found to be uncorrelated with the linear sizes and
opening angles of the radio sources.
While richer environments might be expected to provide greater confining 
pressures and bending forces, the complications introduced by host locations with respect to the cluster center and galaxy velocities obscure any underlying trends. Finally radio power and richness are uncorrelated (Figure 8).
This is consistent with the survey of Ledlow \& Owen (1995a) which found no
correlation between the total radio power of all galaxies in a cluster and
the cluster richness.  Most of our sources occur alone in their clusters
at the sensitivity of {\it FIRST}, and if there are other radio sources in the 
clusters, the bent-double is almost always the dominant source of radio 
power for the cluster.

\section{Correlation with the ROSAT All Sky Survey}
Correlation of the $z<0.5$ complete sample of bent doubles with the 
ROSAT All Sky Survey (RASS) Bright and Faint Source catalogs
reveals that eight of the radio sources lie within one megaparsec of the 
position
of an X-ray source.  In addition to these eight, the two radio sources in 
Coma also
match with detections in the survey.  Matched source \#249 is a quasar at
$z=0.987$, and the X-ray source matched with the bent-double \#082 is 
probably unrelated to the radio source.  This X-ray source is point-like and
there is a 17th magnitude stellar object located $8^{\prime\prime}$ away as
well as a 10th magnitude star located $14^{\prime\prime}$ away.  The X-ray
source is likely associated with one of these objects.
The eight objects (excluding Coma)
are listed in Table 4.  Column (1) lists the bent-double id \#, column
(2) gives the redshift, column (3) shows the match distance between the radio and 
X-ray source in arcminutes, column (4) gives the RASS id name, column (5) lists the 
RASS count rate,
column (6) gives the source extent as listed in the RASS catalog, where a value of zero
indicates that a source is point-like (however, see comments below),
column (7) gives
two estimates for unabsorbed X-ray flux, using a thermal bremsstrahlung model, 
Galactic $N_{H}$ and
temperatures of 3 keV and 1 keV (except for
bent-doubles \#249 and \#082 where no estimates are given) , 
column (8) converts the fluxes in the previous 
column to luminosities, and column (9) gives the optical richnesses for the fields.

When total source counts are low, as is the case for most of our sources, the RASS 
measurement of extent is subject to error.
Therefore, for each of the sources catalogued as point-like, we visually examined the 
RASS images after smoothing them with a $3\sigma$ Gaussian.  This inspection confirmed 
that the X-ray sources associated with objects \#249 and \#082 are point-like, 
while revealing that the source
associated with \#261 could be either point-like or extended, and the X-ray sources
associated with \#186 and \#228 are clearly extended.

Excluding the quasar and the unassociated object, based on our richness 
measurements, five of the six X-ray-detected sources are classified as being 
members of clusters.  The remaining one is listed in Abell's catalog as
a richness class 1 cluster, but falls just short of our cluster
criteria.
One of the other matched objects is also a member of an Abell cluster, and
one is found in a Zwicky cluster.  The sources have typical estimated X-ray 
luminosities of a few times $10^{43}$ erg s$^{-1}$.   A plot of redshift vs. X-ray 
luminosity is given in Figure 9.
The long-dashed lines mark $L_X$ = 10$^{43}$, 10$^{44}$, and 10$^{45}$ erg s$^{-1}$.
The curved lines are the approximate limits of the RASS catalog with the upper 
line (long-short dash) for a  3 keV 
model and the lower (short dash) for a 1 keV model.  These were generated  by taking the
Faint RASS requirement that a source have 6 counts to be in the catalog,
combined with the average exposure for the matching objects of 400 sec,
to get a limiting count rate of 0.015 ct/s, and converting to flux with
PIMMS.  

Above the curves, the solid circles represent the matched sources'
luminosities expected for a 3 keV thermal bremsstrahlung model, using 
Galactic $N_H$, and distances with $q_{\circ}=0.5$ and 
$H_{\circ}=50$ km s$^{-1}$
Mpc$^{-1}$.  The open circles are the same but for a 1 keV model.  
The matched object at $z\approx0.25$ is slightly below
the limit, but was detected because the exposure for that pointing was 
slightly longer (418 s) than average and it had only six counts.
At $L_{X}=10^{42}$, the bent doubles that were not detected in the RASS, but are
found in clusters, based on our richness measurements, are plotted as filled triangles.
Those that are classified as not belonging to clusters and are not detected in the 
RASS, are plotted as asterisks.

Not including Coma, for $z < 0.2$, there are six bent doubles that are members of clusters 
(based on
our richness measurements) and four of these
are detected in the RASS.
We classify nine bent doubles at $z < 0.2$ as non-cluster members, and only
one of these is detected -- the one that is a member
of an Abell cluster and fell just shy of our limit for being called a cluster.  
The majority of the low redshift bent doubles that are found in clusters are 
detected in the RASS, and those that are not found in clusters are not detected 
in the RASS.
This result supports our richness measurements, and suggests that the X-ray flux is probably from cluster, rather than AGN,
emission.

\section{ Conclusions}
We have presented observations of a complete, magnitude-limited sample of 40 
objects taken from our larger
sample of bent-double radio sources.  Our richness measurements indicate that
they can be used as tracers of high-density environments -- 54\% are associated
with Abell class 0 or richer clusters.  The majority of our sample's lowest redshift 
objects ($z < 0.2$) that we classified as cluster members are detected in the ROSAT
All Sky Survey.
Interestingly, 46\% of the sample bent doubles 
are associated with groups, some of them poor.  The lobes in these environments
may be bent by clumped gas in the vicinity of the radio source.  The galaxies
may have high velocities relative to this gas if they are parts of groups
or clusters in formation.  X-ray observations would shed light on this 
scenario.

Some of the sources in the complete sample exist in previously recognized clusters:  six in Abell (1958, 1989) clusters, including two in Coma, and four in Zwicky (1968) clusters.
Within the area of the complete sample, there are a total of 102 clusters
in the Abell catalog, so our sources ``discovered'' $\sim5\%$ of them, 
suggesting, to the extent the Abell catalog is complete,
that $\sim 5\%$ of clusters house bent-double sources.

Almost all of the radio sources in the complete sample are associated with
the first-ranked galaxy in their cluster or group, consistent
with results from other WAT samples.  In addition, their large linear sizes 
(approximately
100 -- 500 kpc) and powers near the FR I/II break are typical of WAT
sources (O'Donoghue et al. 1993).

Few of the radio and optical properties were found to be correlated.
In particular, no correlations are seen between richness of the host environment
and the linear extent or bending angle of the radio lobes.  The complicated 
interactions between
cluster gas, central engine, and extended lobes, as well as the position 
of the host galaxy in the cluster and its velocity relative to the ICM, make 
it unsurprising that there is
not a one-to-one relationship between either richness and linear size or
richness and bending angle.

Before making the magnitude cut, there were 99 bent doubles in the area
of the complete sample, implying that more than half of the radio sources 
have host magnitudes fainter than $m_{R} = 19$.  Given the narrow 
distribution 
in host optical luminosities (Figure 4), these sources must have redshifts 
greater than 
$z\approx0.4$ or 0.5, and can be used to identify such distant clusters with 
an efficiency of at least 50$\%$ (cf. Blanton et al. 2000).

Several of the objects in our sample are in notably low density environments,
and may be associated with fossil groups or single galaxies.
These types of systems may represent a significant mass component of the
Universe that would be missed by optical surveys.
X-ray observations are planned to study
these objects to determine if there is indeed extended gas
associated with them, or if we must find another explanation for the bending
of the radio lobes.

This material is based upon work supported by the National Science
Foundation under Grants AST-98-02732 and 99-70884.  In addition,
support for this work was provided by the National Aeronautics and Space
Administration through $Chandra$ Award Numbers
GO0-1019X, GO0-1141X, and GO0-1173X,
issued by the $Chandra$ X-ray Observatory Center, which is operated by the
Smithsonian Astrophysical Observatory for and on behalf of NASA under
contract NAS8-39073.  Some support was provided by LTSA NAG5-6548.
Part of the work reported here
was done at the Institute of Geophysics and Planetary Physics, under
the auspices of the U.S. Department of Energy by Lawrence Livermore
National Laboratory under contract No.~W-7405-Eng-48.  The {\it FIRST} survey
has also been supported by grants from the National Geographic Society, IGPP,
and Columbia University.

\clearpage

\begin{deluxetable}{ccccc}
\tablenum{1}
\tablewidth{0pt}
\tablecaption{Observations -- Images \& Spectroscopy}
\tablehead{
\colhead{} & \colhead{Images} & \colhead{} & \colhead{Spectra} 
& \colhead{} \\ 
\colhead{id \#} & \colhead{UT} & \colhead{Exp. (s)} & 
\colhead{UT} & \colhead{Exp. (s)}}

\startdata
082  & 02/13/99 & 3$\times$480 &  \nodata & \nodata \\     
087  & 02/24/98 & 3$\times$300 &  02/23/98 & 1200  \\
092  & 02/04/00 & 1$\times$900 &  02/23/98 & 3600 \\
099  & 02/04/00 & 1$\times$900 &  06/02/97 & 1500 \\
105  & 02/04/00 & 1$\times$900 &  02/16/98 & 600  \\
115  & 02/12/99 & 3$\times$480 &  02/11/99 & 1800 \\
116  & 02/24/98 & 3$\times$300 &  02/23/98 &900   \\
119  & 02/13/99 & 3$\times$480 &  02/10/99 & 1800 \\
124  & 02/14/99 & 3$\times$480 &  02/11/99 & 1800  \\
127  & 02/13/99 & 3$\times$480 &  02/10/99 & 1800  \\
135  & 02/04/00 & 1$\times$900 &  02/23/98 & 1200  \\
147  & 02/04/00 & 1$\times$900 &  02/23/98 & 1800  \\
150  & 02/14/99 & 3$\times$480 &  02/10/99 & 1800  \\
153  & 02/24/98 & 1$\times$300 &  06/02/97 & 900 \\
158  & 02/24/98 & 1$\times$300 &  02/23/98 & 1200 \\
161  & 02/14/99 & 3$\times$480 &  02/10/99 & 1800  \\
167  & 02/13/99 & 3$\times$480 &  12/21/98 & 1200  \\
176  & 02/13/99 & 3$\times$480 &  12/21/98 & 1200  \\
177  & 02/04/00 & 1$\times$900 &  05/25/95 & 900 \\
186  & 02/12/99 & 3$\times$480 &  12/22/98 & 1200  \\
189  & 02/14/99 & 3$\times$480 &  02/11/99 & 1200  \\
195  & 02/04/00 & 1$\times$900 &  02/23/98 & 900   \\
196  & 02/14/99 & 3$\times$480 &  02/10/99 & 1800  \\
198  & 02/13/99 & 3$\times$480 &  12/22/98 & 1200  \\
199  & 02/15/99 & 3$\times$480 &  02/10/99 & 1800  \\
200  & 02/12/99 & 3$\times$480 &  02/09/99 & 1200  \\
204  & 02/12/99 & 3$\times$480 &  02/09/99 & 1200  \\
205  & 02/13/99 & 3$\times$480 &  02/09/99 & 1200  \\
213  & 02/04/00 & 1$\times$900 &  02/23/98 & 1800  \\
214  & 01/01/96 & 3$\times$090 &  12/16/96 & 1320 \\
216  & 02/04/00 & 1$\times$900 &  02/23/98 & 1200  \\
217  & 02/15/99 & 3$\times$480 &  02/11/99 & 1800 \\
228  & 02/04/00 & 1$\times$900 &  02/23/98 & 1500  \\
232  & 02/04/00 & 1$\times$900 &  02/23/98 & 1200  \\
234  & 02/04/00 & 1$\times$900 &  02/23/98 & 1200 \\
236  & \nodata & \nodata & \nodata & \nodata   \\
237  & \nodata & \nodata & \nodata & \nodata   \\
249  & 02/04/00 & 1$\times$900  & 02/10/99 & 1200 \\
261  & 02/12/99 & 3$\times$480  & 02/09/99 & 1800 \\
\enddata
\tablecomments{The observations were taken at the MDM 2.4m telescope
with the Echelle CCD and the R-band filter and the Mk III spectrograph with
a few exceptions.  The spectra for objects 99 and 153 were taken
at the Lick Observatory 3m telescope.  The spectral observation of \#177 was
done at the KPNO 4m telescope.  Object \#214  was imaged at the KPNO 4m 
telescope,
and a spectrum was taken at the Keck II Observatory (Blanton et al. 2000).  
Spectra for objects 167, 176, 186, and 198 were contributed by Jules Halpern.
Where no data is presented (objects \#236 and 237 in the Coma cluster and \#82),
observations were taken from the literature.}

\end{deluxetable}

\clearpage

\begin{deluxetable}{ccccccccccccl}
\tabletypesize\scriptsize
\rotate
\tablewidth{0pt}
\tablenum{2}
\tablecaption{The Complete Bent-Double Sample}
\tablehead{
\colhead{id \#} & \colhead{RA(J2000)} & \colhead{Dec(J2000)} & \colhead{m$_{R}$} &
\colhead{Redshift} &  \colhead{Flux\tablenotemark{a}} & 
\colhead{(N/F)\tablenotemark{b}}
 & \colhead{P$_{1440}$} &\colhead{Angle} & \colhead{Size} & \colhead{Size} &
\colhead{FR} &\colhead{Comments} \\ \colhead{} & \colhead{} & \colhead{} & \colhead{} & \colhead{} 
& \colhead{mJy} &\colhead{}&
\colhead{\rm{$\times10^{25}W Hz^{-1}$}} & \colhead{deg.} & \colhead{arcsec} &
\colhead{kpc} & \colhead{} & \colhead{}}
\startdata
082  & 08 03  16.5 &$+24$ 40 35  &  14.50 &   0.045& 140.2 &3.1&     0.12 &   112 &    85 & 102 & I   & CGCG 118-054 \\         
087  & 08 06  11.2 &$+25$ 31 38  &  16.75 &   0.172& 85.7 &1.0  &    1.14 &    89 &    56 & 212 & I &\\
092  & 08 14  46.0 &$+26$ 15 48  &  17.74 &   0.280& 28.1  &1.2  &   1.02 &   145 &    37 & 195 & I &\\        
099  & 08 18  10.6 &$+23$ 32 11  &  17.88 &   0.246& 52.8  &1.5   &  1.46 &    24 &    80 & 389 & I  &\\
105  & 08 28  39.4 &$+24$ 36 55  &  15.57 &   0.087& 350.9 &1.4   &  1.17 &    84 &    81 & 177 & I   &Zw 0825+2447, 4C +24.17\\
115  & 08 39  51.6 &$+29$ 28 18  &  18.50 &   1.137& 139.8 &1.2 &   94.31 &   155 &    66 & 568 & II & quasar \\
116  & 08 44  33.4 &$+26$ 06 15  &  15.63 &   0.109& 136.2 &1.4  &   0.71 &   143 &    62 & 163 & I  &\\  
119  & 08 45  59.4 &$+22$ 31 09  &  18.86 &   0.373& 26.6  &1.2  &   1.74 &   102 &    58 & 360 & I   &\\
124  & 08 50  40.5 &$+29$ 09 07  &  18.51 &   0.433& 34.4  &1.1  &   3.06 &   102 &    33 & 220 & II &\\
127  & 08 53  31.9 &$+23$ 24 00  &  17.74 &   0.306& 43.3  &1.2  &   1.88 &   146 &    56 & 312 & I   &\\
135  & 09 06  52.1 &$+28$ 55 49  &  16.65 &   0.156& 40.7  &1.3  &   0.45 &    92 &    45 & 159 & I &Zw 0903+2907\\        
147  & 09 35  24.9 &$+23$ 55 03  &  17.82 &   0.264& 165.8 &1.2  &   5.30 &    36 &    91 & 462 & I & Abell 0824 \\
150  & 09 42  09.5 &$+30$ 57 17  &  17.88 &   0.297& 26.5  &1.2  &   1.09 &   123 &    45 & 246 & I/II &\\
152  & 09 46  55.4 &$+30$ 23 23  &  18.45 &   0.337& 36.9  &1.0   &  1.95 &   156 &    30 & 176 & II   & uncertain z \\
153  & 09 50  04.6 &$+29$ 17 27  &  17.77 &   0.251& 45.1  &1.4  &   1.31 &   111 &    38 & 187 & I    &Abell 0876\\
158  & 10 06  23.6 &$+24$ 05 26  &  15.57 &   0.075& 69.1  &1.5 &    0.17 &   91  &   18  &  35 &I     &\\
161  & 10 16  02.4 &$+24$ 00 07  &  17.64 &   0.309& 32.4  &1.0  &   1.44 &   105 &    60 & 336 & I    &\\
167  & 10 29  44.9 &$+25$ 23 11  &  17.07 &   0.237& 230.5 &\nodata& 5.91  &    50 &    99 & 469 & I    &\\
176  & 10 45  16.2 &$+23$ 51 40  &  15.88 &   0.140& 34.1  &1.3 &    0.30  &   103 &    49 & 158 & I    &\\
177  & 10 50  10.9 &$+30$ 39 55  &  18.00 &   0.256& 146.2 &1.0  &   4.40  &   127 &    39 & 197 & II   & Zw 1047+3055 \\
186  & 11 08  12.3 &$+26$ 10 35  &  16.36 &   0.175& 116.0 &1.3 &    1.60  &   128 &    68 & 261 & I    &\\
189  & 11 12  20.7 &$+26$ 01 13 &   18.26&    0.728& 55.0  &1.3 &   14.51  &   138 &    44 & 354 & I/II & quasar \\
195  & 11 20  38.5 &$+29$ 12 34  &  17.12 &   0.240& 86.9  &1.5 &   2.28  &   126 &    120& 573 & I    &\\  
196  & 11 20  48.0 &$+30$ 42 03  &  17.55 &   0.241& 15.1  &2.4  &  0.40  &   135 &    52 & 249 & I    &\\
198  & 11 25  59.8 &$+25$ 28 37  &  16.11 &   0.116& 46.8  &1.2  &  0.28  &    85 &    46 & 128 & II   &Abell 1258\\
199  & 11 26  31.4 &$+26$ 38 04  &  18.12 &   0.341& 89.6  &1.2  &  4.89  &    22 &    30 & 178 & I/II &\\
200  & 11 30  48.8 &$+25$ 24 35  &  16.06 &   0.145& 91.5  &1.2   & 0.86  &    74 &    93 & 309 & II   &\\
204  & 11 46  39.9 &$+22$ 41 35  &  17.11 &   0.177& 34.4  &\nodata & 0.49 &    69 &    49 & 190 & I    &\\
205  & 11 48  33.1 &$+23$ 32 26  &  16.87 &   0.176& 34.7  &1.3 &     0.49  &   124 &    38 & 147 & I    &\\
213  & 12 09  38.4 &$+24$ 20 38  &  17.38 &   0.240& 35.3  &1.1 &     0.93  &   83  &   33  & 158 &I     &Zw 1207+2439\\        
214  & 12 09  50.1 &$+28$ 48 07  &  18.90 &   0.336& 15.0  &1.2  &   0.79  &    91 &    30 & 176 & I    &\\
216  & 12 12  58.0 &$+25$ 09 26  &  18.04 &   0.397& 68.2  &1.2  &   5.09  &   138 &    44 & 282 & II   &\\
217  & 12 16  27.4 &$+29$ 28 44  &  18.55 &   0.429& 38.8  &1.4  &   3.40  &   165 &    72 & 479 & I/II & \\
228  & 12 41  08.7 &$+27$ 34 47  &  16.52 &   0.199& 177.6 &1.2  &   3.18  &   126 &    57 & 240 & I    &Abell 1587 \\
232  & 12 46  51.9 &$+30$ 05 21  &  16.24 &   0.140& 125.1 &1.0  &   1.09  &    53 &    63 & 203 & I    &\\
234  & 12 49  42.2 &$+30$ 38 38  &  16.72 &   0.194& 94.7  &1.4  &   1.61  &   104 &    89 & 367 & I    &\\
236  & 12 59  23.3 &$+27$ 54 44 & 14.00\tablenotemark{c}& 0.023& 249.3 &1.5&     0.06 &   135 &    89 &  56 & I    & Abell 1656 (Coma), NGC4869 \\  
237  & 12 59  35.6 &$+27$ 57 35  &  12.28\tablenotemark{d}& 0.023& 206.7 &1.1&     0.05 &    93 &    39 &  25 & I/II & Abell 1656 (Coma), NGC4874 \\
249  & 13 29  03.2 &$+25$ 31 10 &  17.73& 0.987& 89.3  &1.3&    44.72 &   102 &    36 & 306 & II   & quasar, RGB J1329+255A \\
261  & 13 51  03.4 &$+30$ 54 05&  16.77& 0.247&  123.0 &1.1&     3.44 &   176 &    44 & 214 & I/II  &  Zw 1348+3109
\tablenotetext{a}{{\it NVSS} total flux density. $^{b}$Ratio of {\it NVSS} to {\it FIRST} flux density. $^{c}$Colina \& Perez-Fournon (1990). $^{d}$Postman \& Lauer (1995)}
\enddata
\end{deluxetable}
 
\clearpage

\begin{deluxetable}{cccccllcl}
\tabletypesize\scriptsize
\tablenum{3}
\tablewidth{0pt}
\tablecaption{Richness Measurements for the Complete Sample}
\tablehead{
\colhead{id $\#$} & \colhead{Rank} &\colhead{M$_{V}$} & 
\colhead{$m_{R,-19}$} &
\colhead{N$_{0.5,Raw}^{-19}$} & 
\colhead{Bgnd} & 
\colhead{N$_{0.5}^{-19}$}& \colhead{Abell est.}&
\colhead{Notes} \\
\colhead{(1)} & \colhead{(2)} & \colhead{(3)} & \colhead{(4)} & \colhead{(5)} & 
\colhead{(6)} & \colhead{(7)} & \colhead{(8)} & \colhead{(9)}
} 
\startdata
082 &1&   -21.92 &  \nodata&\nodata&\nodata&\nodata&\nodata&poor group\tablenotemark{1}\\
087 &1&   -22.78 &20.53&  22 &     13.9 &  $~8.1\pm 8.4$ & $<0$&\\
092 &1&   -23.04 &21.78&  33 &     18 &       $15\pm10.0$ & 0&\\
099 &1&   -22.60 &21.48&  30 &    19.1 &    $10.9\pm 9.8$ & $<0$&\\
105 &1&   -22.37 &18.94&  25 &    10.6 &    $14.4\pm 8.3$ & $<0$& Zw 0825+2447\\
116 &1&   -22.85 &19.48&  22 &     11.6 &   $10.4\pm 8.1$ & $<0$&\\
119 &1&   -22.68 &22.54&  42 &    24 &        $18\pm11.4$ & 0&\\
124 &1&   -23.57 &23.08&  67 &     29 &       $38\pm13.6$& 1&\\
127 &1&   -23.28 &22.02&  22 &     28 &      $-6\pm10.0$ & $<0$&\\
135 &1&   -22.66 &20.31&  29 &     13.2 &   $15.8\pm 9.0$ & 0& Zw 0903+2907\\
147 &1&   -22.80 &21.62&  58 &     17 &       $41\pm11.7$ & 1&Abell 0824, Rich 0\\
150 &4&   -23.01 &21.89&  44 &     26 &       $18\pm11.7$ & 0&\\
152 &1&   -22.76 &22.21&  33 &    30 &       $~3\pm11.2$ & $<0$&\\
153 &2&   -22.68 &21.45&  47 &     18.1 &   $28.9\pm11.1$ & 1&Abell 0876, Rich 1\\
158 &1&   -22.05 &18.62&  16 &     10.3 &  $~5.7\pm 7.2$ & $<0$&\\
161 &1&   -23.38 &22.02&  38 &     19 &       $19\pm10.5$ & 0 &\\
167 &2&   -23.25 &21.32&  33 &     17.4 &   $15.8\pm 9.9$ & 0&\\
176 &1&   -23.16 &20.04&  25 &     12.5 &   $12.5\pm 8.5$ & $<0$&\\
177 &2&   -22.53 &21.53&  54 &     28 &       $26\pm12.6$ & 0&Zw 1047+3055\\
186 &1&   -23.20 &20.56&  29 &     13.8 &   $15.2\pm 9.1$ & 0&\\
195 &3&   -23.22 &21.35&  25 &     17.5 &  $~7.5\pm 9.2$ & $<0$&\\
196 &1&   -22.79 &21.34&  22 &     17.3 &  $~4.7\pm 8.8$ & $<0$&\\
198 &2&   -22.45 &19.56&  25 &     11.1 &   $13.9\pm 8.3$ & $<0$&Abell 1258, Rich 1\\
199 &1&   -23.15 &22.27&  31 &     24 &      $~7\pm10.5$ & $<0$&\\
200 &1&   -23.02 &20.08&  22 &     12.2 &  $~9.8\pm 8.2$ & $<0$&\\
204 &1&   -22.46 &20.57&  29 &     13.7 &   $15.3\pm 9.1$ & 0&\\
205 &1&   -22.69 &20.56&  44 &     13.7 &   $30.3\pm10.3$ & 1&\\
213 &1&   -23.00 &21.38&  51 &     18.0 &   $33.0\pm11.4$ & 1& Zw 1207+2439\\
214 &1&   -22.24 &22.14&  28 &    14 &        $14\pm 9.0$ & $<0$&\\
216 &1&   -23.69 &22.73&   23 &     31 &     $-8\pm10.4$ & $<0$&\\
217 &1&   -23.41 &22.96&   34 &     13 &      $21\pm 9.4$ & 0&\\
228 &1&   -23.29 &20.81&   40 &     14.3 &  $25.7\pm10.1$ & 0&Abell 1587, Rich 0\\
232 &1&   -22.74 &19.98&   31 &     11.9 &  $19.1\pm 9.0$ & 0&\\
234 &1&   -23.08 &20.80&   15  &    14.7 & $~0.3\pm 7.7$ & $<0$&\\
236&&     -20.85 & \nodata&\nodata&\nodata&\nodata&\nodata&Abell 1656, Rich 2\\
237&2&    -22.57 & \nodata&\nodata&\nodata&\nodata&\nodata&Abell 1656, Rich 2\\
261 &1&   -23.64 &21.41& 41 &     17.8 &     $23.2\pm10.6$ & 0&\\
\enddata
\tablecomments{
Column (1) lists the bent double i.d. numbers, column (2) shows the rank
of the galaxy's optical magnitude within its group or cluster,
column (3) lists the absolute V-magnitude of the radio source host galaxy,
column (4) gives the apparent R-magnitude corresponding to $M_{V} = -19$ at
the redshift of the radio source,
column (5) lists the number of galaxies within a radius of 0.5 Mpc from the radio 
galaxy, and with magnitudes brighter than M$_{V} = -19$,
column (6) lists the field galaxy counts taken directly from the frame for those
bent doubles that have $m_{R,-19} > 21.5$ or from Metcalfe et al. (1991)
for those that have $m_{R,-19} < 21.5$, column (7) gives the
richness value (raw counts - background (field) counts) along with the Poisson
error,
column (8) lists our estimate of the Abell richness class based on the
N$_{0.5}$ counts, and column (9) lists known cluster associations.}
\tablenotetext{1}{Stocke \& Burns (1987)}
\end{deluxetable}

\begin{deluxetable}{ccccccccl}
\tablenum{4}
\tabletypesize\footnotesize
\tablewidth{0pt}
\tablecaption{Correlation with the ROSAT All Sky Survey}
\tablehead{
\colhead{id \#} & \colhead{z} & \colhead{d\tablenotemark{a}} & \colhead{RASS id} & \colhead{Ct. Rate} & \colhead{Extent\tablenotemark{b}} &
\colhead{$F_{X}$\tablenotemark{c}} & \colhead{$L_{X}$\tablenotemark{d}} & \colhead{N$_{0.5}^{-19}$\tablenotemark{e}} \\
\colhead{} & \colhead{} & \colhead{($\prime$)} & \colhead{} & \colhead{(ct/s)} & \colhead{(arcsec)} &
\colhead{($\times 10^{-13}$)} & \colhead{($\times 10^{43}$)} & \colhead{}}

\startdata

249 & 0.987 & 0.18 & 1RXS J132903.8+253103 & 0.04559 & 0 &\nodata & \nodata & quasar \\
205 & 0.176 & 0.49 & 1RXS J114834.8+233208 & 0.02600 & 23 &  4.03/3.66  &  4.23/3.84    &	30.3     \\
261 & 0.247 & 0.55 & 1RXS J135103.8+305332 & 0.01449 & 0 &  1.87/1.64  &  3.53/3.09    &	23.2 Zw	 \\
186 & 0.175 & 0.66 & 1RXS J110811.8+260956 & 0.03642 & 0 &  5.03/4.46  &  5.21/4.62    &	15.2   	 \\
198 & 0.116 & 0.90 & 1RXS J112602.8+252801 & 0.02426 & 32 &  3.29/2.91  &  1.62/1.43    &	13.9 A 	 \\
232 & 0.140 & 1.94 & 1RXS J124659.1+300411 & 0.01793 & 11 &  2.32/2.03  &  1.60/1.40    &	19.1   	 \\
228 & 0.199 & 1.29 & 1RXS J124104.2+273535 & 0.07328 & 0 &  8.39/7.16 &   6.48/5.53   &	25.7 A 	 \\
082 & 0.045 & 5.42 & 1RXS J080307.8+243532 & 0.03276 & 0 &  \nodata & \nodata &	poor group   	 \\
\tablenotetext{a}{Match distance, in arcminutes, between the radio source and RASS source}
\tablenotetext{b}{Source extent, as listed in the RASS catalog, where a value 
of zero indicates a point source.  However, visual examination of the RASS 
images associated with sources with a catalogued value of zero
confirms that the X-ray sources matching with \#249 and \#082 are point 
sources, the source associated with \#261 could be point-like or extended, and
the sources associated with \#186 and \#228 are clearly extended.}
\tablenotetext{c}{Unabsorbed X-ray flux in units of 10$^{-13}$ erg cm$^{-2}$ s$^{-1}$, 
in the $0.12-2.48$ keV band, assuming a thermal bremsstrahlung model, Galactic $N_{H}$, 
and kT = 3 keV / 1 keV.}
\tablenotetext{d}{X-ray luminosity in units of 10$^{43}$ erg s$^{-1}$, in 
the $0.12-2.48$ keV band, using the flux from the previous column and assuming 
$H_{\circ} = 50$ km s$^{-1}$ Mpc$^{-1}$ and $q_{\circ} = 0.5$.}
\tablenotetext{e}{Richness value.  A = Abell cluster, Zw = Zwicky cluster.}
\enddata
\end{deluxetable}

\clearpage
\begin{figure}[t!]
\figurenum{1}
\plotone{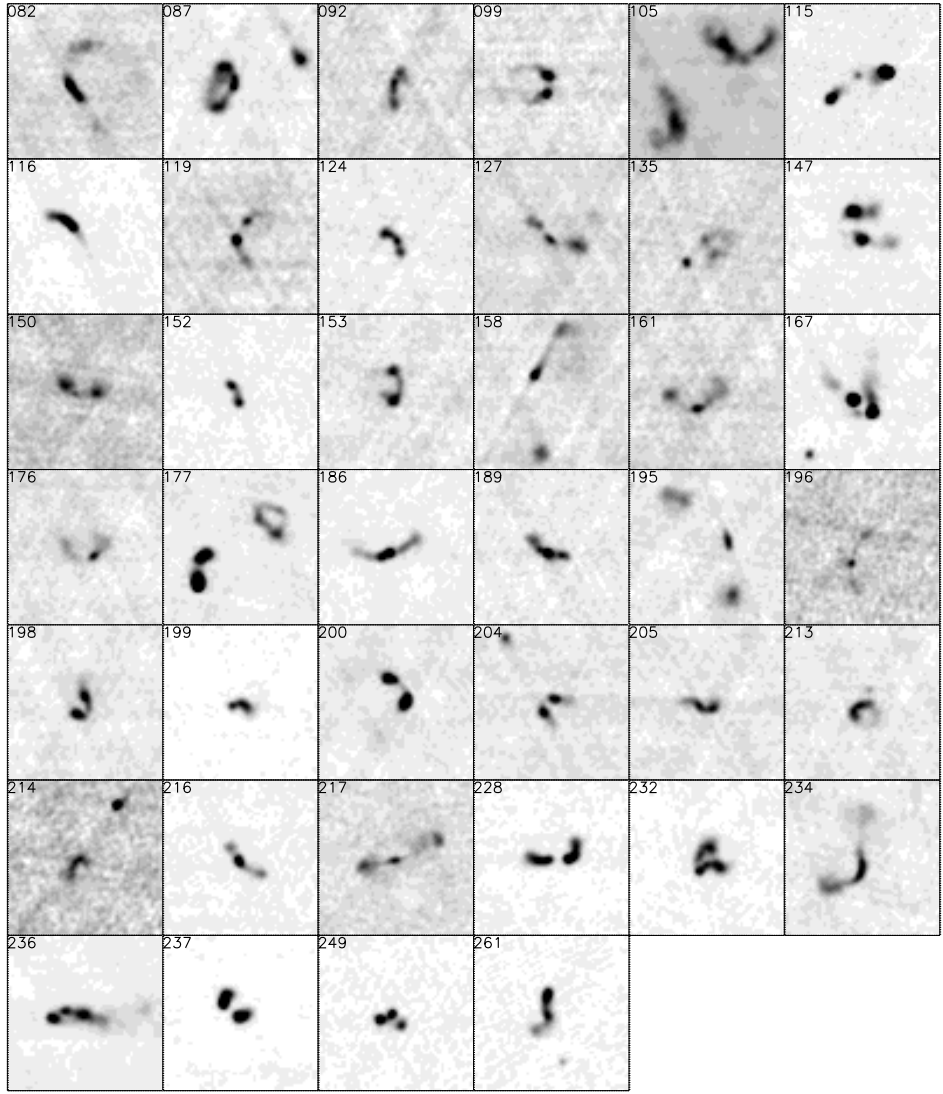}
\caption{Grayscale postage-stamp plots of the complete sample {\it FIRST} 
bent-double radio sources.  Images are squares of $2 \times 2$ arcmin.}
\end{figure}

\clearpage
\begin{figure}[t!]
\figurenum{2}
\plotone{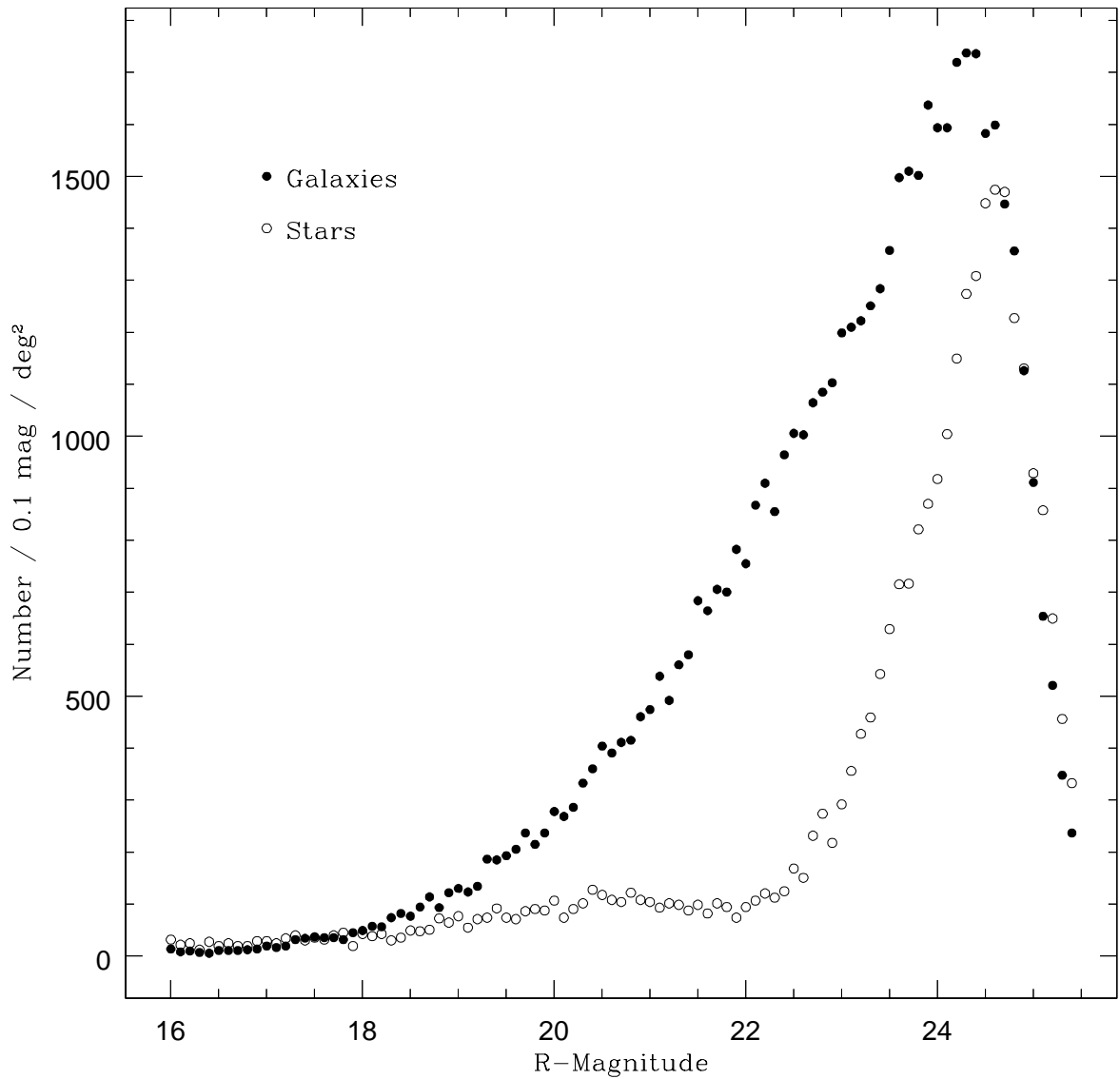}
\caption{Number counts of galaxies and stars as a function of R-magnitude 
derived from the sum of the 33 fields from which we derived richness 
measurements.  The completeness limit is reached at $m_{R} \approx 24$ and the
star/galaxy classification breaks down at $m_{R} \approx 22$.}
\end{figure}

\clearpage
\begin{figure}[t!]
\figurenum{3a}
\plotone{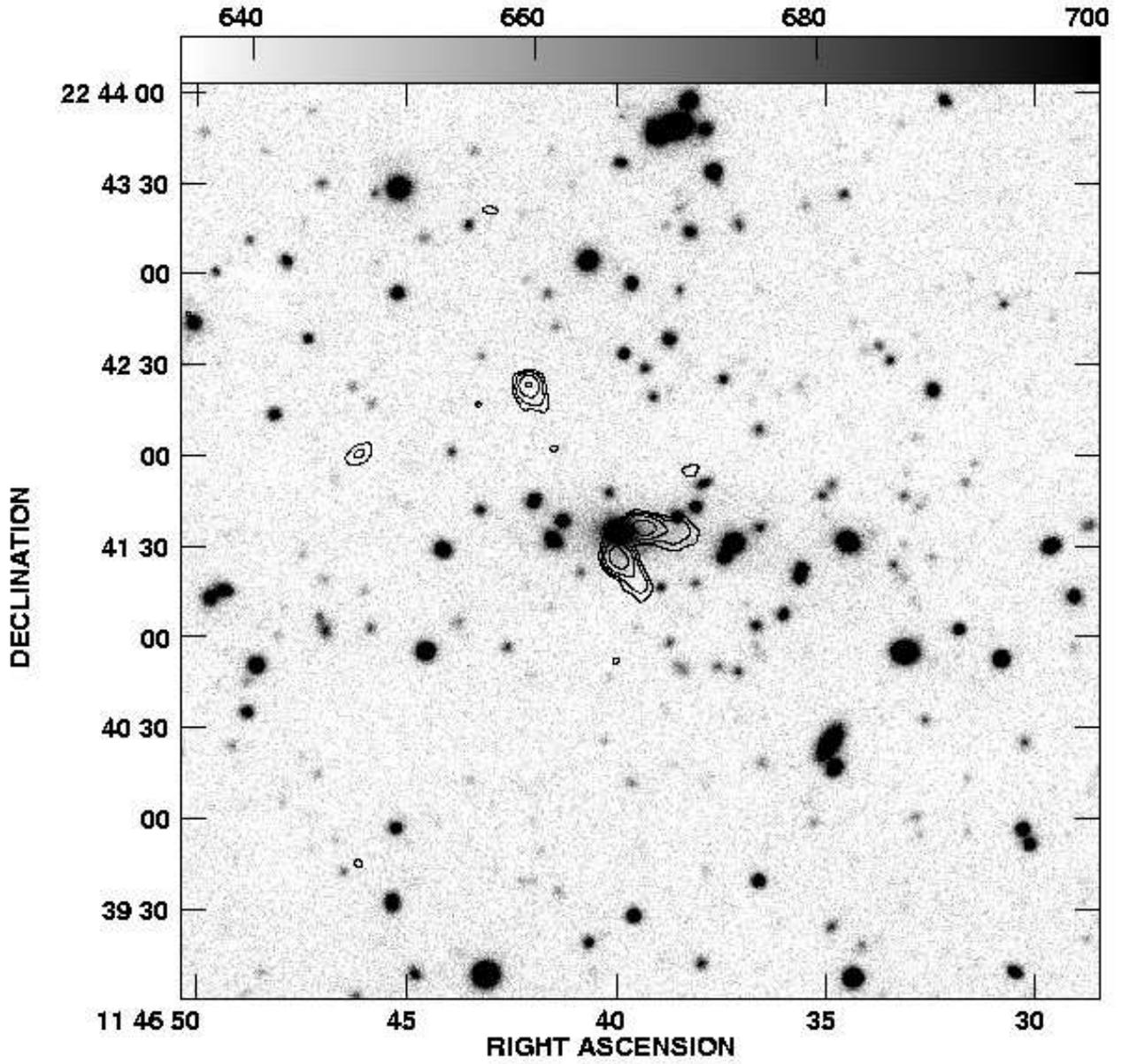}
\caption{A bent double in a rich environment: 1146+2241 (\#204) at $z=0.1773$.}
\end{figure}

\clearpage
\begin{figure}[t!]
\figurenum{3b}
\plotone{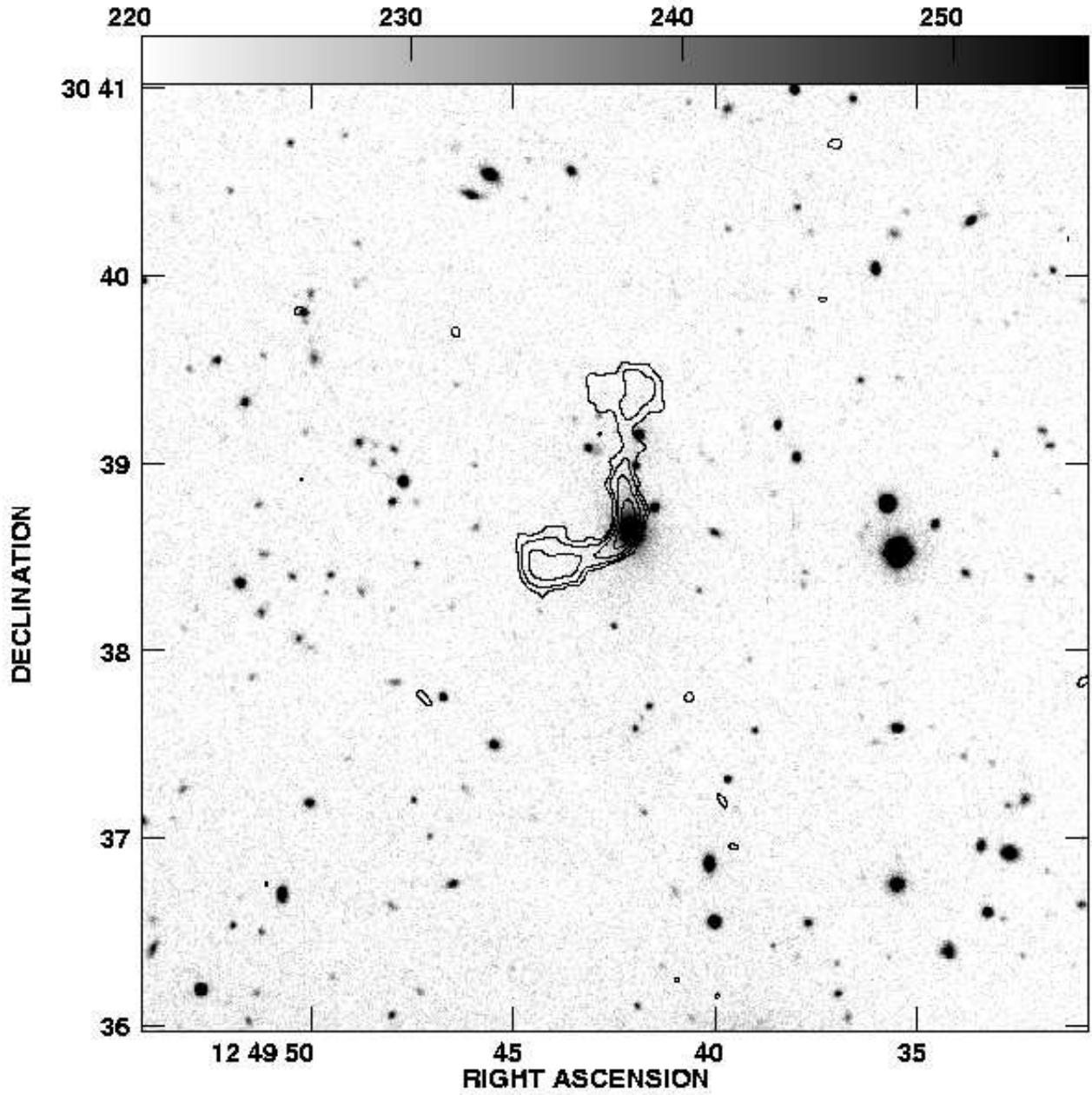}
\caption[Poor environment, 1249+3038]{A bent double in a poor 
environment:1249+3038 (\#234) at $z=0.1935$.}
\end{figure}

\clearpage
\begin{figure}[t!]
\figurenum{4}
\plotone{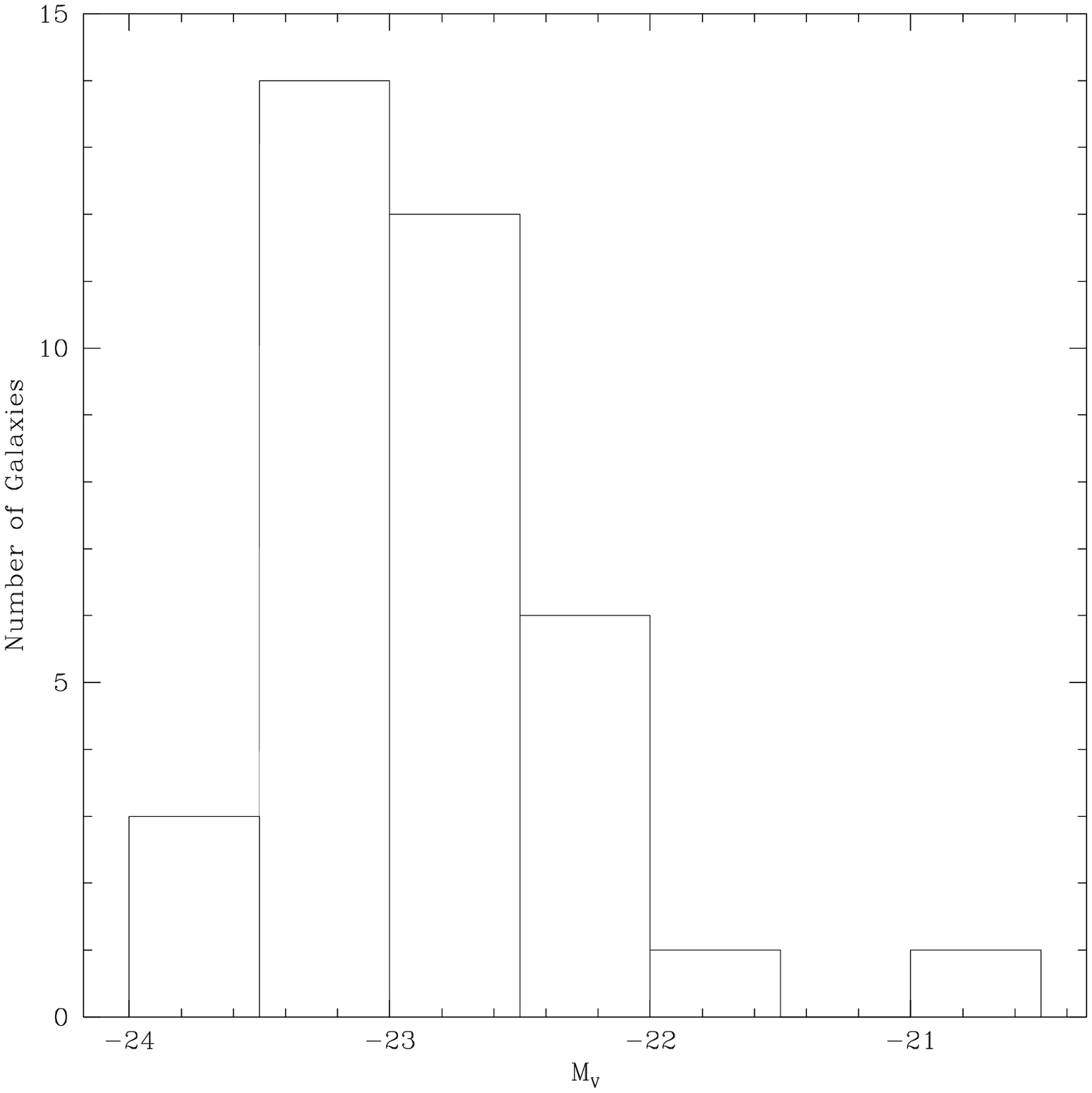}
\caption[Redshift vs. R-magnitude]{A histogram of the distribution of
absolute magnitudes for the bent-double host galaxies.}
\end{figure}

\clearpage
\begin{figure}[t!]
\figurenum{5}
\plotone{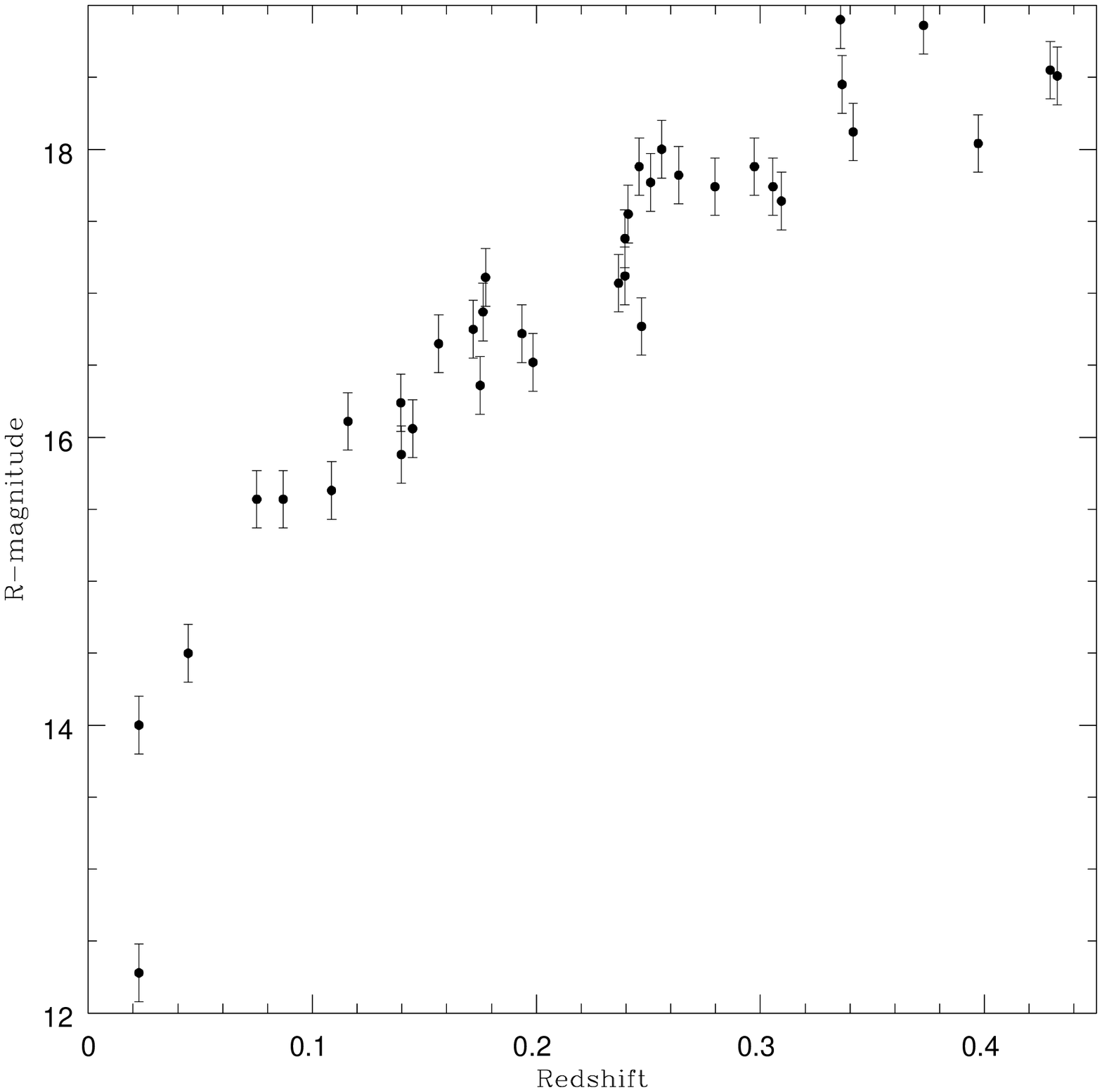}
\caption{Redshift vs. R-magnitude for the complete sample of
bent doubles.  Error bars are $\pm0.2$ mag.  All but 8 of the 40 sources are the 1st ranked galaxy in
their cluster or group.  The distribution looks very similar to that expected
from no-evolution elliptical galaxy models (Coleman, Wu, \& Weedman 1980).}
\end{figure}

\clearpage
\begin{figure}[t!]
\figurenum{6}
\plotone{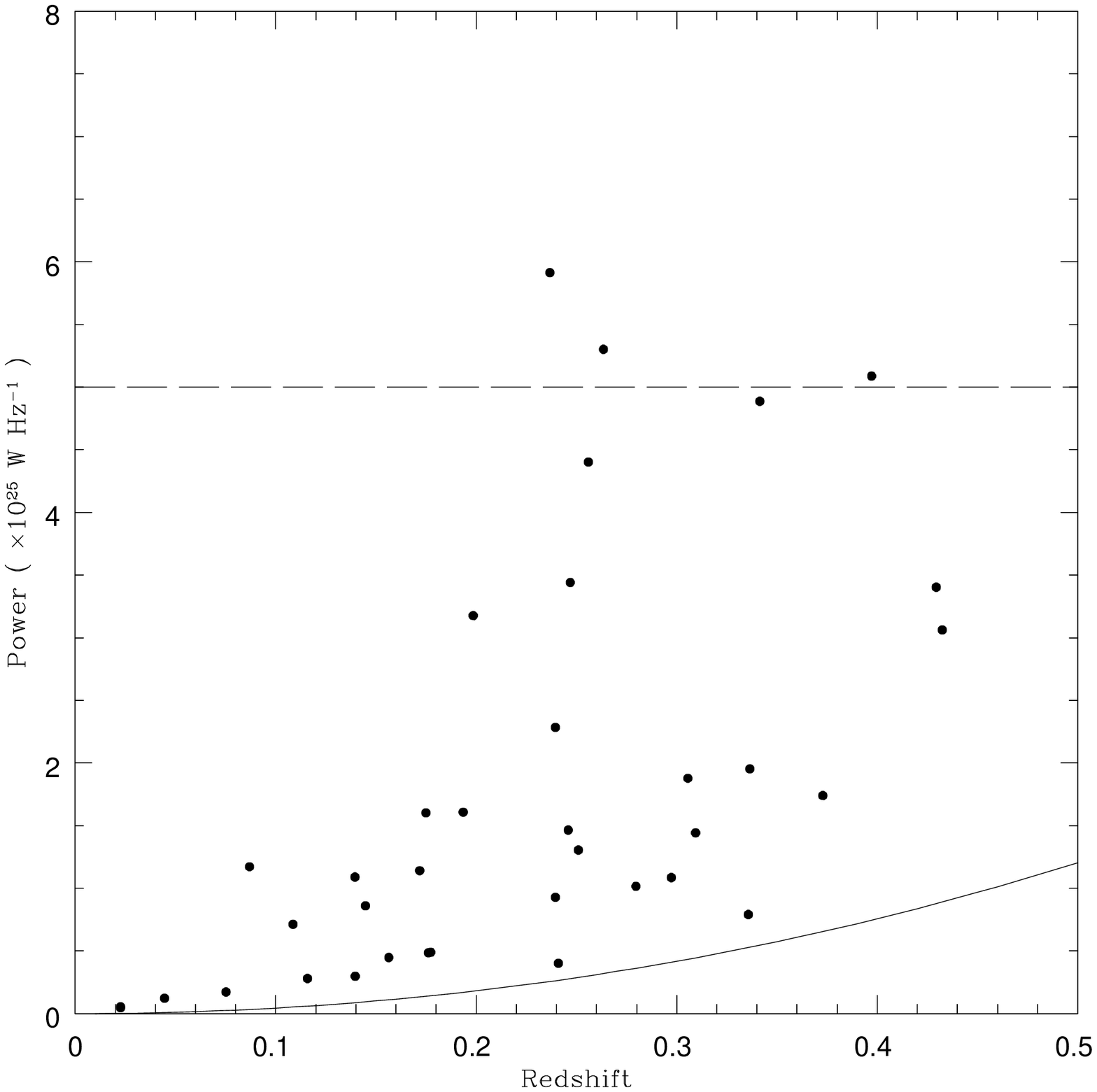}
\caption[Radio power vs. redshift]{The apparent increase in radio power 
with redshift
is probably due to selection effects.  Faint objects at high-$z$ will be
missed because of the flux limit at which we can recognize a bent double. 
The objects in the complete sample have fluxes greater
than 10 mJy (the solid line on the plot).  The dashed line shows the 
approximate boundary in power between FR I and II sources -- most of the 
bent doubles fall below the line and are FR I's.}
\end{figure}

\clearpage
\begin{figure}[t!]
\figurenum{7}
\plotone{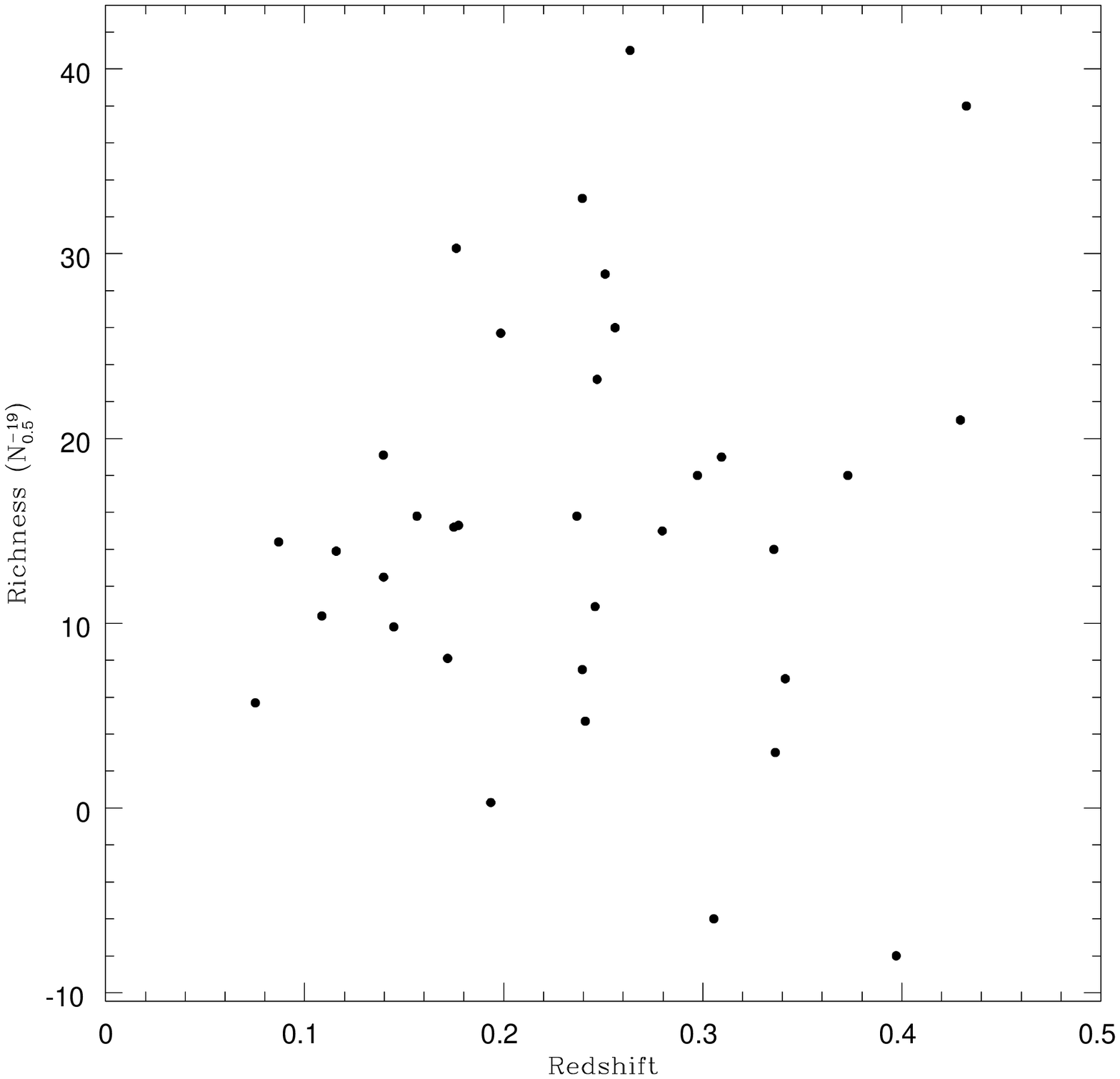}
\caption[Richness vs. redshift]{The redshift of the radio host galaxy and the
richness of its associated cluster are not correlated.  Bent doubles are found in clusters up to $z \approx 0.5$ that are as rich as those at the present day.}
\end{figure}

\clearpage
\begin{figure}[t!]
\figurenum{8}
\plotone{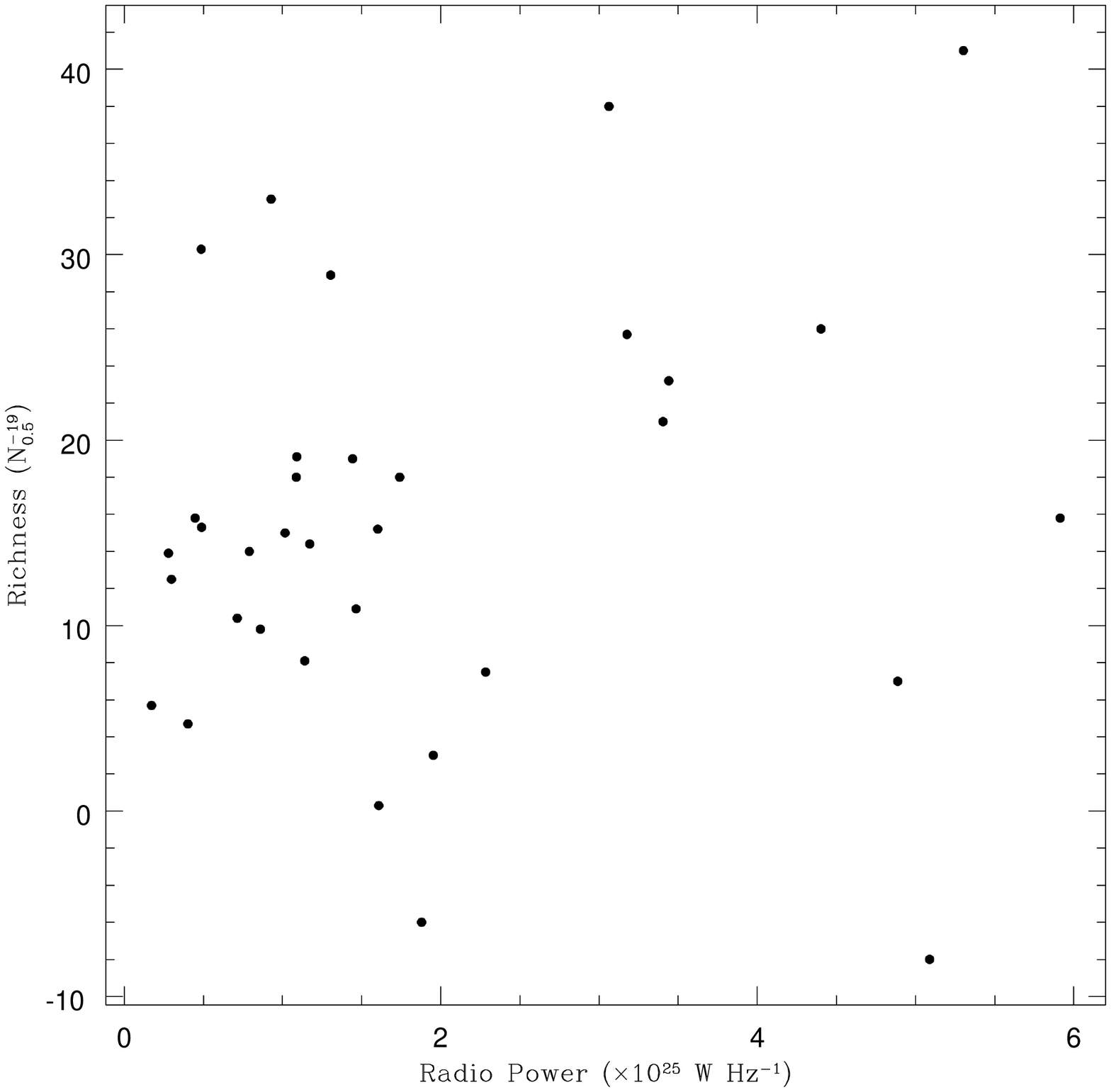}
\caption[Richness vs. radio power]{Radio power and cluster richness are not
correlated.  This is consistent with results from the VLA survey of Abell clusters by Ledlow \& Owen (1995a).}
\end{figure}

\clearpage
\begin{figure}[t!]
\figurenum{9}
\plotone{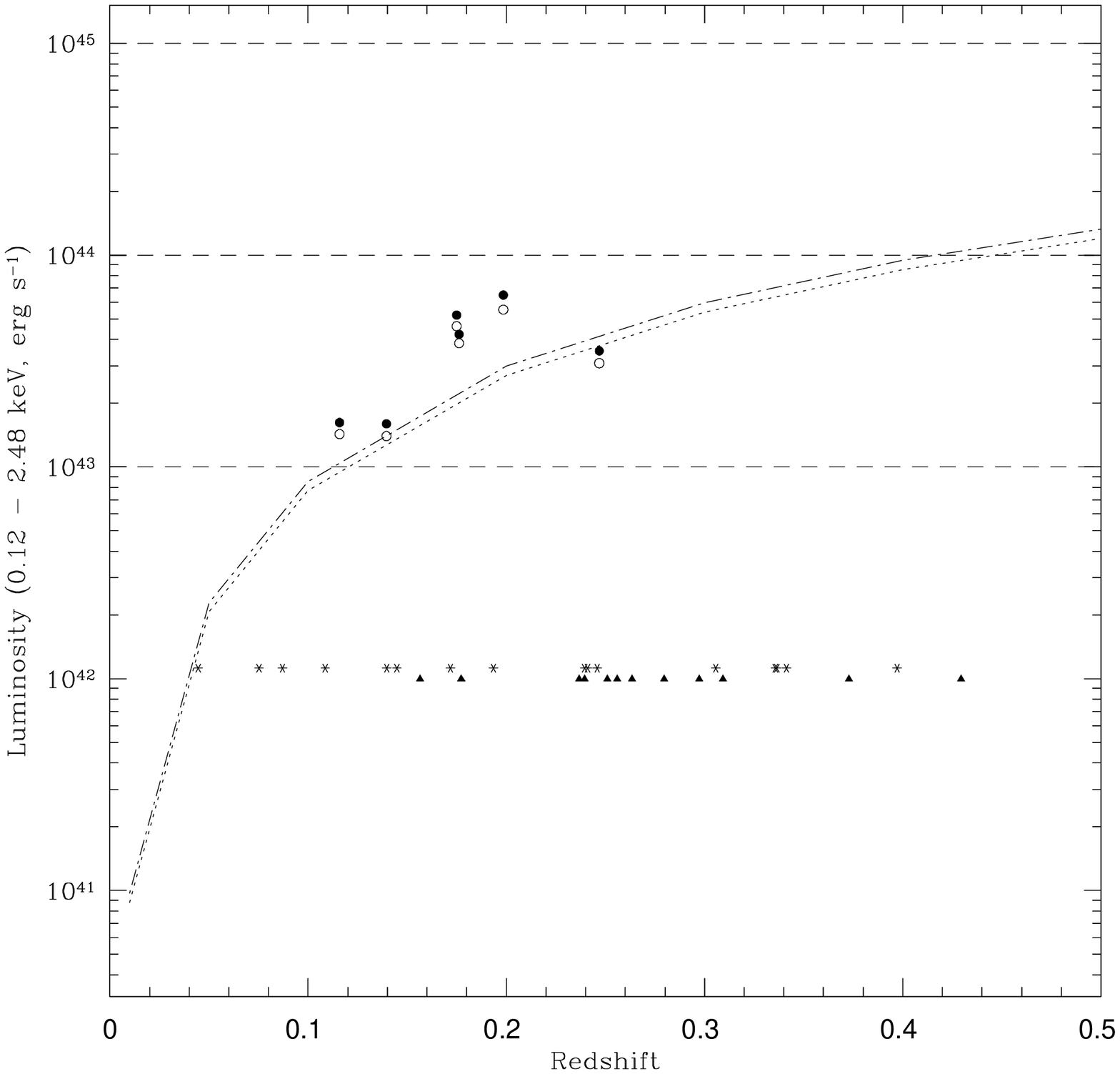}
\caption[]{Curved lines are the approximate limits of the RASS catalog with 
the upper 
line (long-short dash) for a  3 keV 
and the lower (short dash) for a 1 keV thermal bremsstrahlung model.
Above the curves, the solid/open circles represent the detected source
luminosities expected for the 3/1 keV model.
At $L_{X}=10^{42}$, the bent doubles that were not detected in the RASS, 
but are found in clusters are plotted as filled triangles. Those that are 
classified as not belonging 
to clusters and are not detected in the RASS, are plotted as asterisks.}
\end{figure}


\begin{references}

\reference{} Abell, G. O. 1958, \apjs, 3, 211                                    

\reference{} Abell, G. O., Corwin, H. G., Jr., \& Olowin, R. P. 1989, ApJS, 70, 1

\reference{} Allington-Smith, J. R., Ellis, R. S., Zirbel, E. L., \& Oemler, A. 
1993, \apj,404, 521                                                                   

\reference{} Becker, R. H., White, R. L., \& Helfand, D. J. 1995, \apj, 450, 559

\reference{} Blanton, E. L., Gregg, M. D., Helfand, D. J., Becker, R. H., \& White, R. L. 2000, \apj, 531, 118

\reference{} Burns, J. O., Rhee, G., Roettiger, K., \& Owen, F. 1993, in 
Observational Cosmology,                                     
eds. G. Chincarini, A. Iovino, T. Maccacaro, \& D. Maccagni (ASP Conf. Ser., 51),
 407                 

\reference{} Burns, J. O., Rhee, G., Owen, F. N., \& Pinkney, J. 1994, \apj, 423,
 94
                                
\reference{} Burns, J. O., G\'omez, P., Pinkney, J. Roettiger, K., \& Loken, C. 
1996,                                                 
in Clusters, Lensing, and the Future of the Universe, eds. V. Trimble \& A. 
Reisenegger (ASP Conf. Ser., 88), 184                     


\reference{} Coleman, G. D., Wu, C.-C., \&  Weedman, D. W. 1980, ApJS, 43, 393

\reference{} Colina, L. \& Perez-Fournon 1990, ApJS, 72, 41

\reference{} Condon, J. J., Cotton, W. D., Greisen, E. W., Yin, Q. F., Perley, R.
 A., Taylor, G. B., \& Broderick, J. J. 1998, \aj, 115, 1693


\reference{} Deltorn, J.-M., Le F\`evre, O., Crampton, D., \& Dickinson, M. 1997,
 483 L21                    

\reference{} Dickinson, M. 1997, in The Early Universe with the VLT,
ed. J. Bergeron, Berlin:Springer, 274


\reference{} Fanaroff, B. L., \& Riley, J. M. 1974, \mnras, 167, 31L

\reference{} G\'omez, P. L., Pinkney, J., Burns, J. O., Wang, Q., Owen, F. N., 
Vo
ges, W. 1997, \apj, 474, 580

\reference{} Hill, G. J., \& Lilly, S. J. 1991, \apj, 367, 1

\reference{} Landolt, A. U. 1992, \aj, 104, 340

\reference{} Ledlow, M. J. \& Owen, F. N. 1995a, \aj, 109, 853

\reference{} Ledlow, M. J. \& Owen, F. N. 1995b, \aj, 110, 1959

\reference{} Ledlow, M. J. \& Owen, F. N. 1996, \aj, 112, 9

\reference{} Ledlow, M. J., Owen, F. N., \& Eilek, J. A. 2000, astro-ph/9908336

\reference{} Lilly, S. J., Cowie, L. L., \& Gardner, J. P., \apj, 369, 79

\reference{} Metcalfe, N., Shanks, T., Fong, R., \& Jones, L. R.  1991, \mnras, 2
49, 498

\reference{} O'Donoghue, A. A., Eilek, J. A., \& Owen, F. N. 1993, \apj, 408, 428
                                                     
\reference{} Owen, F. N., \& Rudnick, L. 1976, \apj, 205, L1

\reference{} Owen, F. N. \& White, R. A. 1991, \mnras, 249, 164

\reference{} Ponman, T. J et al. 1994, Nature, 369, 462

\reference{} Postman, M. \& Lauer, T R. 1995, \apj, 440, 28

\reference{} Rector, T. A., Stocke, J. T., Ellingson, E. 1995, \aj, 110, 1492

\reference{} Roettiger, K., Burns, J. O., \& Loken, C. 1996, \apj, 473, 651

\reference{} Romer, A. K. et al. 2000, ApJS, 126, 209

\reference{} Schlegel, D. J., Finkbeiner, D. P., \& Davis, M. 1998, 500, 525

\reference{} Steidel, C. C. \& Hamilton, D.  1993, \aj, 105, 2017

\reference{} Stocke, J. T. \& Burns, J. O. 1987, \apj, 319, 671


\reference{} Stocke, J. T., Perlman, E. S., Gioia, I. M., \& Harvanek, M. 1999,
\aj, 117, 1967

\reference{} Tonry, J. \& Davis, M. 1979, \aj, 84, 10

\reference{} Valdes, F. 1983, \it{Faint Object Classification and Analysis System
}, \rm{Central Computer Services, National Optical Astronomy Observatories, P.O. 
Box 26732, Tucson, AZ  85725}

\reference{} Vikhlinin, A. et al. 1999, \apj, 520, L1

\reference{} Zirbel, E. L. 1997, \apj, 476, 489

\reference{} Zwicky, F., Herzog, E., Wild, P., Karpowicz, M., \& Kowal, C. T. 
1968, Catalogue of                                      
Galaxies and Clusters of Galaxies, (California Institute of Technology, Pasadena)
                                                     

\end{references}
\end{document}